\documentclass[12pt,preprint]{aastex}
\usepackage{amssymb}
\usepackage{graphicx}

\begin{document}
\title{Hyperaccreting Neutron-Star Disks and Neutrino Annihilation}
\author{Dong Zhang and Z. G. Dai}
\affil{Department of Astronomy, Nanjing University, Nanjing 210093,
China; \\\rm dongzhanghz@gmail.com, dzg@nju.edu.cn}

\begin{abstract}
Newborn neutron stars surrounded by hyperaccreting and
neutrino-cooled disks may exist in some gamma-ray bursts (GRBs)
and/or supernovae (SNe). In this paper we further study the
structure of such a neutron-star disk based on the two-region (i.e.,
inner \& outer) disk scenario following our previous work, and
calculate the neutrino annihilation luminosity from the disk in
various cases. We investigate the effects of the viscosity parameter
$\alpha$, energy parameter $\varepsilon$ (measuring the neutrino
cooling efficiency of the inner disk) and outflow strength on the
structure of the entire disk as well as the effect of emission from
the neutron star surface boundary emission on the total neutrino
annihilation rate. The inner disk satisfies the entropy-conservation
self-similar structure for the energy parameter $\varepsilon\simeq
1$ and the advection-dominated structure for $\varepsilon<1$. An
outflow from the disk decreases the density and pressure but
increases the thickness of the disk. Moreover, compared with the
black-hole disk, the neutrino annihilation luminosity above the
neutron-star disk is higher, and the neutrino emission from the
boundary layer could increase the neutrino annihilation luminosity
by about one order of magnitude higher than the disk without
boundary emission. The neutron-star disk with the
advection-dominated inner disk could produce the highest neutrino
luminosity while the disk with an outflow has the lowest. Although a
heavily mass-loaded outflow from the neutron star surface at early
times of neutron star formation prevents the outflow material from
being accelerated to a high bulk Lorentz factor, an energetic
ultrarelativistic jet via neutrino annihilation can be produced
above the stellar polar region at late times if the disk accretion
rate and the neutrino emission luminosity from the surface boundary
layer are sufficiently high.
\end{abstract}


\section{Introduction}
The hyperaccreting disk surrounding a stellar-mass black hole
possibly formed by the merger of a compact object binary or the
collapse of a massive star has been argued to be a candidate for
central engines of gamma-ray bursts (GRBs) (e.g. Eichler et al.
1989; Narayan et al. 1992; Woosley 1993; Paczy\'{n}ski 1998; Popham
et al. 1999; MacFadyen \& Woosley 1999; Narayan et al. 2001). The
typical mass of the debris dense torus or disk is about $0.01-1
M_{\odot}$ with large angular momentum as well as high accretion
rate up to $\sim 1.0 M_{\odot}\,{\rm s}^{-1}$. Although the optical
depth of the accreting matter in the disk is enormous, the disk can
be cooled or partly cooled via neutrino emission. A number of
studies have investigated the structure and energy transfer of the
neutrino-cooled disk around a black hole both in steady-state and
time-dependent considerations over last several years (Popham et al.
1999; Narayan et al. 2001; Kohri \& Mineshige 2002; Di Matteo et al.
2002; Kohri et al. 2005; Gu et al. 2006; Kawanaka \& Mineshige 2007;
Chen \& Beloborodov 2007; Liu et al. 2007; Janiuk et al. 2007;
Metzger et al. 2008).

An alternative model of central engines of GRBs is newly, rapidly
rotating neutron stars or magnetars (Usov 1992; Klu\'{z}niak \&
Ruderman 1998; Dai \& Lu 1998; Ruderman et al. 2000; Wheeler et al.
2000). In recent years, newborn neutron stars have also been
suggested as an origin of some GRBs and their afterglows. For
example, Dai et al. (2006) argued that the X-ray flares discovered
by {\em Swift} can be explained as being due to magnetic instability
and reconnection-driven events from highly-magnetized millisecond
pulsars; the shallow decay phase of X-ray afterglows is considered
to be due to energy injection to a forward shock by a relativistic
pulsar wind (Dai 2004; Yu \& Dai 2007); a newly-formed neutron star
rather than a black hole is expected to explain the light curve of
SN 2006aj associated with GRB 060218 (Mazzali et al. 2006; Soderberg
et al. 2006). Moreover, simulations on the merger of a compact
object binary show that it is possible to form a hypermassive
neutron star, at least a transiently existing neutron star after the
merger, depending on initial conditions of the binary, equations of
state of neutron matter and the structure of magnetic fields
(Shibata et al. 2003; Shibata 2003; Lee \& Ramirez-Ruiz 2007;
Anderson et al. 2008; Liu et al. 2008). Therefore, the
hyperaccreting disk around a neutron star can also be considered as
possible central engines for some GRBs. Based on these motivations,
we have studied the structure of the hyperaccretion disk around a
neutron star using both analytic and numerical methods (Zhang \& Dai
2008, hereafter ZD08). We found that the neutron-star disk can cool
more efficiently and produce a higher neutrino luminosity than the
black-hole disk.

In ZD08, the quasi-steady disk around a neutron star is
approximately divided into two regions --- inner and outer disks,
depending on the energy transfer and emission in the disk. For the
outer disk, the heating energy rate $Q^{+}$ is mainly due to local
dissipation ($Q^{+}=Q_{\rm vis}^{+}$), and the structure of the
outer disk is very similar to the black-hole disk. On the other
hand, the heating energy in the inner disk includes both the energy
generated by itself and the energy advected from the outer region
($Q^{+}=Q_{\rm vis}^{+}+Q_{\rm adv}^{+}$), so the inner disk has to
be dense with a high pressure. We approximately take $Q^{+}=Q^{-}$
and the entropy-conservation self-similar condition $ds$=0 to
describe the inner disk. The size of the inner disk is determined by
the global energy equation of the inner disk. However, we need to
point out that the entropy-conversation structure is not the only
possible structure of the inner disk, which depends on the detailed
form of energy and mass transfer. In the case where $Q^{-}<Q^{+}$ in
the inner disk, we should take the advection-dominated self-similar
structure to describe the inner disk.

The net gravitational binding energy of the accreting matter is
proposed to drive a relativistic outflow or jet by two general
mechanisms that could provide energy for GRBs: neutrino annihilation
and magnetohydrodynamical mechanisms such as the Blandford-Znajek
effect. The mechanism of neutrino annihilation is easy to understand
and could be calculated based on the structure and neutrino
luminosity in the disk (Ruffert et al. 1997, 1998; Popham et al.
1999; Asano \& Fukuyama 2000, 2001; Di Matteo et al. 2002; Miller et
al. 2003; Birkl et al. 2007; Gu et al. 2006; Liu et al. 2007).
However, the annihilation rate due to neutrino emission from the
black-hole disk may not be able to produce a sufficiently high
luminosity to explain some energetic GRBs (Di Matteo et al. 2002).
Gu et al. (2006) and Liu et al. (2007) showed that the annihilation
luminosity can reach $10^{52}$ergs s$^{-1}$ even for an accretion
rate $\sim 10M_{\odot}$ s$^{-1}$. However, such an accretion rate is
too large for energetic long GRBs, since this requires an
unreasonable massive accretion disk around a compact object. As the
neutron-star disk structure and neutrino luminosity are different
from the black-hole disk, it is interesting to calculate the
neutrino annihilation rate above the neutron-star disk and to
consider whether the annihilation energy rate and luminosity above a
neutron-star disk are high enough to produce energetic GRBs.

On the other hand, we do not consider any outflow from the disk and
neutron star in ZD08, which may play a significant role in the
structure and energy transfer in the disks around neutron stars. A
nonrelativistic or mildly relativistic ouflow or ``wind" from the
disk can be considered as an energy source of supernovae (MacFadyen
\& Woosley 1999; Kohri et al. 2005). This theoretical model becomes
more attractive after the discovery of the connection between some
GRBs and supernovae (e.g. Galama et al. 1998, Stanek et al. 2003,
Prochaska et al. 2004, Campana et al. 2006), while the GRB component
is considered from a relativistic jet produced by neutrino
annihilation. As a result, both outflow ejection and neutrino
annihilation could be important in the events of GRB-SN connections
within the framework of the collaspar model (Woosley \& Bloom 2006).
However, an outflow from the black-hole disk is expected and becomes
important whenever the accretion flow is an advection-dominated
accretion flow (ADAF) (Narayan \& Yi 1994, 1995), while the neutrino
luminosity is relatively low for ADAF. In other words, neutrino
emission may not provide a sufficiently high amount of energy for
GRBs associated with supernovae if a thermally-driven outflow is
produced from the advection-dominated disk at the same time.
Therefore, we need to calculate the neutrino luminosity and
annihilation efficiency of the advection-dominated disk with outflow
around a neutron star if the neutrino luminosity is much higher for
the advection-dominated neutron-star disk than the black-hole disk.

In this paper we still consider the case in which the central object
is a neutron star rather than a black hole. Our purpose is to
further study the structure of a hyperaccreting neutron-star disk
following ZD08, and calculate the neutrino annihilation rate above
such a disk. This paper is organized as follows. In \S2 we introduce
basic equations of the neutrino-cooled disk. We discuss the
properties of the inner disk in \S3 based on the two-region disk
scenario introduced in ZD08. We study the disk with different values
of the viscosity parameter $\alpha$ and the energy parameter
$\varepsilon$ (some quantities are given in this paper in Table 1),
and then study the disk structure with an outflow. Two models of an
outflow driven from the neutron-star disk are introduced in \S3.4.
In \S4, we calculate the neutrino annihilation rate and luminosity
above the neutron-star disk in various cases, and compare the
results with the black hole disk. We discuss the effect of the
neutrino luminosity at the neutron star surface boundary layer on
the annihilation rate. In \S5, we particularly focus on an
astrophysical application of the neutron-star disk in GRBs and
GRB-SN connections. Conclusions are presented in \S6.

\section{Basic Equations}
\subsection{Conservation equations}
In this paper, all quantities are used as their usual meanings (see
ZD08). We adopt the cylindrical coordinates ($r, \varphi, z$) to
describe the disk. $v_{r}$, $v_{\varphi}$ are the radial and
rotational velocity, $\Omega$ is the angular velocity, and
$\Omega_{K}$ is the Keplerian angular velocity. $\Sigma=2\rho H$ is
the disk surface density with $\rho$ as the density and $H$ as the
half-thickness of the disk. Vertical hydrostatic equilibrium gives
$H=c_{s}/\Omega_{K}$, where the isothermal sound speed is
$c_{s}=(P/\rho)^{1/2}$ with $P$ to be the gas pressure. The
$\nu_{k}=\alpha c_{s}H$ is the kinematic viscosity coefficient in
the disk with $\alpha$ to be the viscosity parameter.

The mass continuity equation is
\begin{equation}
\frac{1}{r}\frac{d}{dr}\left(r\Sigma
v_{r}\right)=2\dot{\rho}H,\label{a01}
\end{equation}
where $\dot{\rho}$ is the mass-loss term. If the outflow of the disk
is weak, the mass accretion rate $\dot{M}$ can be considered as a
constant and we have the accretion rate,
\begin{equation}
\dot{M}=-4\pi r\rho v_{r}H\equiv -2\pi r\Sigma v_r.\label{a02}
\end{equation}
In \S 3.3 we will also discuss the disk structure with outflows.

The angular momentum conservation reads
\begin{equation}
\Sigma rv_{r}\frac{d(rv_{\varphi})}{dr}=\frac{d}{dr}\left(\Sigma
\alpha
\frac{c_{s}^{2}}{\Omega_{K}}r^{3}\frac{d\Omega}{dr}\right)+\frac{d}{dr}\dot{J}_{ext},\label{a03}
\end{equation}
with $\dot{J}_{ext}$ as the external torque acted on the disk, such
as the torque acted by the outflow from the disk. The angular
momentum flows into the central compact star or the coupling exerted
by the star on the inner edge of the disk is
$C=-\dot{M}(GMr_{*})^{1/2}$ with $r_{*}$ being the neutron star
radius (Frank et al. 2002). Therefore, for a weak outflow, combined
with equation (\ref{a01}) and the above boundary condition, equation
(\ref{a03}) is integrated as
\begin{equation}
\nu
\Sigma=\frac{\dot{M}}{3\pi}\left(1-\sqrt{\frac{r_{*}}{r}}\right),\label{a04}
\end{equation}
where $r_{*}$ is the  neutron star radius. Here we adopted the
standard assumption that the torque is zero at the inner boundary of
the disk $r_{*}+b$ with $b\ll r_{*}$. In \S 3.3 we will discuss the
angular momentum equation with outflow.

The energy equation of the disk is
\begin{equation}
\Sigma v_{r}T\frac{ds}{dr}=Q^{+}-Q^{-},\label{a05}
\end{equation}
where $T$ is the temperature in the disk and $s$ is the local
entropy per unit mass, $Q^{+}$ and $Q^{-}$ are the heating and
cooling energy rates in the disk. In the outer disk, the energy
input is mainly due to the local viscous dissipation,
\begin{equation}
Q^{+}=Q^{+}_{\rm vis}=\frac{3GM\dot{M}}{8\pi
r^{3}}\left(1-\sqrt{\frac{r_{*}}{r}}\right).\label{a06}
\end{equation}
The left term of equation (\ref{a05}) can be taken as the energy
advection term $Q_{\rm adv}^{-}$. We can obtain (ZD08)
\begin{equation}
Q_{\rm adv}^{-}=\Sigma
v_{r}T\frac{ds}{dr}=v_{r}T\frac{\Sigma}{2r}\left[\frac{R}{2}\left(1+Y_{e}\right)+\frac{4}{3}g_{*}\frac{aT^{3}}{\rho}
\right],\label{a07}
\end{equation}
where $R=8.315\times10^{7}$ergs mole$^{-1}$ K$^{-1}$ is the gas
constant, $Y_{e}$ is the ratio of electron to nucleon number
density, the free degree factor $g_{*}$ is 2 for photons and 11/2
for a plasma of photons and relativistic $e^{-}e^{+}$ pairs.

The energy cooling rate $Q^{-}$ is mainly due to neutrino emission,
i.e., $Q^{-}\approx Q^{-}_{\nu}$ (Popham \& Narayan 1995; Di Matteo
et al. 2002),
\begin{equation}
Q_{\nu}^{-}=\sum_{i=e,\mu,\tau}\frac{(7/8)\sigma_{B}T^{4}}{(3/4)[\tau_{\nu_{i}}/2+1/\sqrt{3}+1/(3\tau_{a,\nu_{i}})]}
.\label{a08}
\end{equation}
The three types of neutrino cooling rate per unit volume are
\begin{equation}
\dot{q}_{\nu_{e}}=\dot{q}_{\rm eN}+\dot{q}_{e^{-}e^{+}\rightarrow
\nu_{e}\bar{\nu}_{e}}+\dot{q}_{\rm brem}+\dot{q}_{\rm
plasmon},\label{a081}
\end{equation}
\begin{equation}
\dot{q}_{\nu_{\mu}}=\dot{q}_{\nu_{\tau}}=\dot{q}_{e^{-}e^{+}\rightarrow
\nu_{\tau}\bar{\nu}_{\tau}}+\dot{q}_{\rm brem}, \label{a082}
\end{equation}
where $\dot{q}_{\rm eN}$, $\dot{q}_{e^{-}e^{+}\rightarrow
\nu_{i}\bar{\nu}_{i}}$, $\dot{q}_{\rm brem}$, and $\dot{q}_{\rm
plasmon}$ are the electron-positron pair capture rate, the
electron-positron pair annihilation rate, the nucleon bremsstrahlung
rate, and the plasmon decay rate. Following Kohri et al. (2005),
Janiuk et al. (2007) and Liu et al. (2007), we calculate the
absorption and scattering optical depth for three types of neutrinos
$\tau_{a,\nu_{i}(e,\mu,\tau)}$ and $\tau_{s,\nu_{i}(e,\mu,\tau)}$ as
well as the neutrino cooling rates. For hyperaccretion disks, the
electron-positron pair capture rate plays the most important role
among several types of neutrino cooling rates.

Moreover, besides the local energy equation (\ref{a05}), we need the
global energy conservation equation of the inner disk in order to
decide the size of the inner disk. The maximum power that the inner
disk an release is estimated as (ZD08)
\begin{eqnarray}
L_{\nu,max}&\approx&\frac{3GM\dot{M}}{4}\left\{\frac{1}{3r_{*}}-\frac{1}{r_{\rm
out}}\left[1-\frac{2}{3}\left(\frac{r_{*}}{r_{\rm
out}}\right)^{1/2}\right]\right\}\nonumber\\
&&-\bar{f}_{\nu}\frac{3GM\dot{M}}{4}\left\{\frac{1}{\tilde{r}}
\left[1-\frac{2}{3}\left(\frac{r_{*}}{\tilde{r}}\right)^{1/2}\right]-
\frac{1}{r_{\rm out}}\left[1-\frac{2}{3}\left(\frac{r_{*}}{r_{\rm
out}}\right)^{1/2}\right]\right\}\nonumber\\
&&\approx\frac{3GM\dot{M}}{4} \left\{{\frac{1}{3
r_{*}}-\frac{\bar{f}_\nu}{\tilde{r}}\left[1-\frac{2}{3}
\left(\frac{r_{*}}{\tilde{r}}\right)^{1/2}\right]}\right\},
\end{eqnarray}
where $\tilde{r}$ is the radius between inner and outer disks (i.e.,
the size of the inner disk), the average neutrino cooling efficiency
$\bar{f}_{\nu}$ is determined by
\begin{equation} \bar{f}_{\nu}=\frac{\int_{\tilde{r}}^{r_{\rm
out}}Q_{\nu}^{-}2\pi r dr}{\int_{\tilde{r}}^{r_{\rm out}}Q^{+}2\pi r
dr}.\label{a10}
\end{equation}
Thus we derive
\begin{equation}
\int_{r_{*}}^{\tilde{r}}Q_{\nu}^{-}2\pi rdr=\varepsilon
L_{\nu,max}=\varepsilon \frac{3GM\dot{M}}{4} \left\{{\frac{1}{3
r_{*}}-\frac{\bar{f}_\nu}{\tilde{r}}\left[1-\frac{2}{3}
\left(\frac{r_{*}}{\tilde{r}}\right)^{1/2}\right]}\right\}\label{a09}
\end{equation}
with the energy parameter $\varepsilon$ being introduced to measure
the neutrino cooling efficiency of the inner disk. When the outer
disk flow is mainly an ADAF, we have $\bar{f}_\nu\sim 0$ and the
maximum energy release rate of the inner disk to be
$GM\dot{M}/4r_{*}$. When the outer disk flows is an efficiently
NDAF, then $\bar{f}_\nu\simeq 1$ and the energy release of the inner
disk mainly results from the heat energy generated by itself. The
values of $\bar{f}_\nu$ are calculated analytically in Zhang (2009,
Fig. 3.6). In ZD08, we simply set $\varepsilon=1$ and use the
entropy-conservation self-similar structure to describe the inner
disk. In \S3, we also discuss the case of $\varepsilon<1$ with
different structures of the inner disk. In addition, if we consider
an outflow ejected from the disk, equation (\ref{a09}) should be
modified. We will discuss the modification in \S3.3.

\subsection{Pressure and $\beta$-equilibrium}
The total pressure in the disk is the summation of four terms:
nucleons, radiation, electrons (including $e^{+}e^{-}$ pairs) and
neutrinos,
\begin{equation}
P=P_{\rm nuc}+P_{\rm rad}+P_{e}+P_{\nu},\label{b01}
\end{equation}
where the pressures of nucleons, radiation and electrons are
\begin{equation}
P_{\rm nuc}=\frac{\rho k_{B}T}{m_{B}},\label{b02}
\end{equation}

\begin{equation}
P_{\rm rad}=\frac{1}{3}aT^{4},\label{b03}
\end{equation}

\begin{equation}
P_{e^{\pm}}=\frac{1}{3}\frac{m_{e}^{4}c^{5}}{\pi^{2}\hbar^{3}}
\int^{\infty}_{0}\frac{x^{4}}{\sqrt{x^{2}+1}}\frac{dx}{e^{(m_{e}c^{2}\sqrt{x^{2}+1}\mp\mu_{e})/k_{B}T}+1},\label{b05}
\end{equation}
and
\begin{equation}
P_{e}=P_{e^{-}}+P_{e^{+}}.\label{b04}
\end{equation}
We use the Fermi-Dirac distribution to calculate the pressure of
electrons, where $\mu_{e}$ is the chemical potential of electron
gas, and $k_{B}$ is the Boltzmann constant.

The ratio of the neutrino pressure to the total pressure becomes
noticeable only in very opaque regions of the disk (e.g., Kohri et
al. 2005, their Fig. 6; Liu et al. 2007, their Fig. 3). The neutrino
pressure is
\begin{equation}
P_{\nu}=u_{\nu}/3, \label{b06}
\end{equation}
where $u_{\nu}$ is the energy density of neutrinos. We adopt the
expression of $u_{\nu}$ from previous work (e.g., Di Matteo et al.
2002).

The adiabatic index of the accreting matter is important to
determine the size of the inner disk when it satisfies the
entropy-conservation condition. It can be written as
\begin{equation}
\gamma=1+(P_{\rm nuc}+P_{\rm rad}+P_{e}+P_{\nu})/(u_{\rm nuc}+u_{\rm
rad}+u_{e}+u_{\nu}). \label{b07}
\end{equation}

Moreover, we need the equation of charge neutrality
\begin{equation}
n_{p}=\frac{\rho Y_{e}}{m_{B}}=n_{e^{-}}-n_{e^{+}},\label{b08}
\end{equation}
and the chemical equilibrium equation
\begin{eqnarray}
n_{p}(\Gamma_{p+e^{-}\rightarrow
n+\nu_{e}}+\Gamma_{p+\bar{\nu}_{e}\rightarrow
n+e^{+}}+\Gamma_{p+e^{-}+\bar{\nu}_{e}\rightarrow n}) \nonumber\\=
n_{n}(\Gamma_{n+e^{+}\rightarrow
p+\bar{\nu}_{e}}+\Gamma_{n\rightarrow
p+e^{-}+\nu_{e}}+\Gamma_{n+\nu_{e}\rightarrow p+e^{-}})\label{b09}
\end{eqnarray}
to determine the matter components in the disk, where $n_{p}$,
$n_{e^{-}}$ and $n_{e^{+}}$ are the number densities of protons,
electrons and positrons, and the various weak interaction rates
$\Gamma_{p\rightarrow n}$ ($\Gamma_{n\rightarrow p}$) can be
calculated following Janiuk et al. (2007; see also Kawanaka \&
Mineshige 2007). When neutrinos are perfectly thermalized, we derive
the $\beta$-equilibrium distribution in the disk
\begin{equation}
\textrm{ln}\left(\frac{n_{n}}{n_{p}}\right)
=f(\tau_{\nu})\frac{2\mu_{e}-Q}{k_{B}T}+[1-f(\tau_{\nu})]\frac{\mu_{e}-Q}{k_{B}T},\label{b10}
\end{equation}
with $Q=(m_{n}-m_{p})c^{2}$, and the factor $f(\tau_{\nu})=\rm
exp(-\tau_{\nu_{e}})$ combines the formula from the
neutrino-transparent limit case with the the neutrino-opaque limit
case of the $\beta$-equilibrium distribution. However, we should
keep in mind that the $\beta$-equilibrium is established only if the
neutronization timescale $t_{n}$ is much shorter than the accretion
timescale $t_{a}$ in the disk $t_{n}<t_{a}$. Beloborodov (2003)
found that the equilibrium requires the accretion rate $\dot{M}$ to
satisfy
\begin{equation}
\dot{M}>\dot{M}_{eq}=2.24\times10^{-3}(r/10^{6}\textrm{cm})^{13/10}(\alpha/0.1)^{9/5}
(M/M_{\odot})^{-1/10}M_{\odot}s^{-1}.\label{b11}
\end{equation}
When the accretion rate is sufficiently low, the electron fraction
$Y_{e}$ would freeze out from weak equilibrium, while the disk
becomes advection-dominated (e.g., Metzger et al. 2008b, 2009). In
this case, the chemical composition of the disk is determined by its
initial condition before its evolution. Metzger et al. (2009), for
example, showed that the hyperaccreting disk around a black hole
generically freeze out with the fixed $Y_{e}\sim 0.2-0.4$. We will
discuss the effect of chemical equilibrium on neutron-star disks
more detailedly in the next section. The $\beta$-equilibrium
assumption can be approximately adopted in our calculations even for
the ADAF case.

\section{Properties of the Disk}
The two-region disk scenario in ZD08 allows the gravitational energy
of the neutron-star disk system to be released in three regions:
outer disk, inner disk and neutron-star surface. The inner disk
region is formed due to the prevention effect of the neutron-star
surface, i.e., most advection energy generated in the disk still
need to be released in a region near the neutron-star surface.
Moreover, a difference between the angular velocity  of the
neutron-star surface and that of the disk inner boundary layer leads
to neutrino emission in the surface boundary layer. In ZD08, we
assume all the advected energy to be released in the disk and
furthermore all the advected energy from the outer region to be
released in the inner disk. Actually, it is possible that a part of
the advected energy can be transferred onto the neutron-star surface
and finally cooled by neutrino emission from the surface boundary
layer rather than the inner disk. Local microphysics quantities such
as the inner energy density, neutrino cooling rate, heating
convection and conduction properties as well as the advection and
cooling timescales should be calculated in order to simulate the
inner disk formation and the cooling efficiency of the steady-state
inner disk. In this paper, however, we adopt a simple method to
determine the inner disk structure, i.e., we use the global energy
equation (\ref{a09}) with the global parameter $\varepsilon$ instead
of the local energy equation (\ref{a05}). The inner disk can release
all the advected energy transferred inward for $\varepsilon=1$,
while a part of the advected energy can still be transferred onto
the neutron-star surface for $\varepsilon<1$. Moreover, we
approximately take $\varepsilon$ as a constant in the inner disk,
and adopt the self-similar treatment to calculate the inner disk
structure. We take $Q^{-}=Q^{+}$ or the entropy-conservation
condition $ds=0$ for $\varepsilon=1$ in the entire inner disk, and
$Q^{-}=\varepsilon Q^{+}$ for $\varepsilon<1$ with the
advection-dominated self-similar structure.

With the energy parameter $\varepsilon$ in the global energy
equation (\ref{a09}) and the self-similar treatment, the inner disk
model can be simplified and calculated by assuming the accretion
rate, the mass of the central compact object, and the self-similar
structure of the inner disk. We discussed the disk structure with
the entropy-conservation condition in ZD08, and in this section we
further discuss the properties of the neutron-star disk in various
cases. Furthermore, we need to consider the effect of an outflow
from the neutron-star disk.

\subsection{Entropy-Conservation Inner Disk with Different $\alpha$}
The viscosity parameter $\alpha$ was first used by Shakura \&
Sunyaev (1973) to express the relation between viscous stress
$t_{r\theta}$ and the pressure $P$ in the disk as
$t_{r\theta}=\alpha P$. Another formula introduces the turbulent
kinematic viscosity is $\nu_{k}=\alpha c_{s}H$ (Frank et al. 2002).
MHD instability simulations show a wide range of $\alpha$ from 0.6
to about 0.005 or less (Hawley et al. 1995; Brandenburg et al. 1995;
Balbus \& Hawley 1998; King et al. 2007). King et al. (2007)
summarized observational and theoretical estimates of the disk
viscosity parameter $\alpha$. They showed that there is a large
discrepancy between the typical values of $\alpha$ from the best
observational evidence ($\alpha\sim 0.1-0.4$ for fully ionized thin
disks) and those obtained from numerical MHD simulations
($\alpha\leq 0.02$ and even considerably smaller). More elaborate
numerical simulations should be carried out for resolving this
problem. For neutrino-cooled hyperaccreting disks, many previous
papers choose $\alpha=0.1$ as the most typical value. The disk
structure with $\alpha$ from 0.01 to 0.1 was discussed in Chen \&
Beloborodov (2007). On the other hand, hyperaccreting disks with
very low $\alpha$ have also been discussed. For example, Chevalier
(1996) studied the neutrino-cooled disk with an extremely small
$\alpha\sim 10^{-6}$. In this section, we discuss the neutron-star
disk with different $\alpha$. We choose the value of $\alpha$ from
0.001 to 0.1. The size of the inner disk alters with different
$\alpha$, and we still adopt the entropy-conversation self-similar
condition of the inner disk to study the effects of the viscosity
parameter.

As discussed in ZD08, from equation (\ref{a09}), if
$\varepsilon\simeq 1$, the heating energy advected from the outer
disk together with the energy generated in the inner disk is totally
released in the inner disk, and the energy balance can be
established between heating and cooling in the inner disk, i.e.,
$Q^{+}=Q^{-}$, or from equation (\ref{a05}), $Tds/dr=0$. In the case
where $v_{r}\ll v_{K}$ and $\Omega\sim \Omega_{K}$ but
$|\Omega-\Omega_{K}|\geq v_{r}/r$, we can obtain the
entropy-conservation self-similar structure of the inner disk
(Medvedev \& Narayan 2001; ZD08) by
\begin{equation}
\rho \propto r^{-1/(\gamma-1)},\,\, P \propto
r^{-\gamma/(\gamma-1)},\,\, v_r \propto
r^{(3-2\gamma)/(\gamma-1)}.\label{c01}
\end{equation}
Since the adiabatic index of the accretion matter is not constant,
we modify expression (\ref{c01}) as
\begin{eqnarray}
\frac{\rho(r)}{\rho(r+dr)}=\left(\frac{r}{r+dr}\right)^{-1/(\gamma(r)-1)},
\frac{P(r)}{P(r+dr)}=\left(\frac{r}{r+dr}\right)^{-\gamma(r)/(\gamma(r)-1)},
\nonumber\\\frac{v(r)}{v(r+dr)}=\left(\frac{r}{r+dr}\right)^{(3-2\gamma(r))/(\gamma(r)-1)}.\label{c02}
\end{eqnarray}

The size of the inner disk $\tilde{r}$ with various $\alpha$ can be
determined by equation (\ref{a09}) in \S 2.1. Figure 1 shows the
size of the inner disk as a function of accretion rate with
different viscosity parameter $\alpha$=0.1, 0.01 and 0.001. Same as
in ZD08, the outer edge radius of the inner disk $\tilde{r}$
decreases with increasing the accretion rate for a low accretion
rate when most part of the disk is advection-dominated. $\tilde{r}$
reaches its minimum value at $\dot{M} \sim 0.1 M_{\odot}s^{-1}$ for
$\alpha=0.1$, and then increases with increasing the accretion rate.
However, the size of the inner disk becomes smaller for lower
viscosity parameter $\alpha$ and would expand dramatically for a
higher accretion rate. The characteristic rate $\dot{M}_{0}$ which
minimizes the size of inner disk $\tilde{r}$ can be approximated by
$\dot{M}_{\rm ch}\sim \alpha M_{\odot}$ s$^{-1}$. The value of
accretion rate $\dot{M}_{\rm ch}$ is between those of characteristic
rates $\dot{M}_{\rm ign}$ (rate of ignition) and $\dot{M}_{\rm
opaque}$ (rate of transparency) in Chen \& Beloborodov (2007).

Actually, from ZD08, we derive an approximate analytic equation of
the radius $\tilde{r}$ between the innner and outer disks as
\begin{equation}
\tilde{r}^{\frac{5(5-3\gamma)}{4(\gamma-1)}}\left(1-\sqrt{\frac{r_{*}}{\tilde{r}}}\right)^{3/4}
\left(r_{*}^{\frac{3\gamma-8}{2(\gamma-1)}}-
\tilde{r}^{\frac{3\gamma-8}{2(\gamma-1)}}\right) \propto
\dot{M}^{-3/2}\alpha_{-1}^{5/2}\label{c03}
\end{equation}
for a radiation-pressure-dominated ADAF outer disk, and
\begin{equation}
\left(1-\sqrt{\frac{r_{*}}{\tilde{r}}}\right)^{-31/11}\tilde{r}^{(\frac{-1}
{\gamma-1}+\frac{3}{22})}\left(r_{*}^{\frac{2-3\gamma}{\gamma-1}}-
\tilde{r}^{\frac{2-3\gamma}{\gamma-1}}\right)^{-1}
\propto\dot{M}^{10/11}\alpha_{-1}^{-21/11}\label{c04}
\end{equation}
for a gas-pressure-dominated ADAF. In both cases, the inner disk
size delines as the accretion rate increases or the viscosity
parameter $\alpha$ decreases. For a neutrino-dominated disk, the
inner disk size $\tilde{r}$ reaches its minimum value
\begin{equation}
\left(\frac{\gamma-1}{3\gamma-2}\right)
\tilde{r}^{\frac{2\gamma-1}{\gamma-1}}
\left(1-\sqrt{\frac{r_{*}}{\tilde{r}}}\right)\left(r_{*}^{\frac{2-3\gamma}{\gamma-1}}-
\tilde{r}^{\frac{2-3\gamma}{\gamma-1}}\right)
\propto\left\{\frac{1}{r_{*}}-\frac{3\bar{f}_{\nu}}{\tilde{r}}\left[1-\frac{2}{3}
\left(\frac{r_{*}}{\tilde{r}}\right)^{1/2}\right]\right\}.\label{c05}
\end{equation}
The solution of minimum $\tilde{r}$ declines as $\alpha$ decreases
because the lower-$\alpha$ disk has a higher value of
$\bar{f}_{\nu}$. As a result, the inner disk size $\tilde{r}$ always
decrease with decreasing $\alpha$ in the ADAF case. This conclusion
is consistent with that of Figure 1.

For simplicity, we here adopt an unified model introduced in \S 2 to
calculate the structure of the disk both in the ADAF and NDAF cases.
In particular, we assume that the disk is always in the
$\beta$-equilibrium state. Let us focus on this equilibrium
assumption. Following Beloborodov (2003, i.e., equation [\ref{b11}]
in our paper), we plot the characteristic curves of equilibrium with
different values of $\alpha$ in the $\dot{M}-\tilde{r}$ plane of
Figure 1. The $\beta$-equilibrium state can only be established in
the right region divided by the corresponding curve. In the left
region, as mentioned at the end of \S 2.2, the weak interaction
timescale become longer than the disk evolutionary timescale, and
the electron fraction $Y_{e}$ freezes out with a fixed value.
However, based on the analytic and numerical arguments in ZD08, we
found that the solutions of disk structure are relatively
insensitive to the value of $Y_{e}$, and the main results of Figure
1 can still be unchanged for various $Y_{e}$. In ZD08, we fixed the
value of $Y_{e}=1/9$ and $Y_{e}=1$ as the two limits. The inner disk
size increases slightly with increasing $Y_{e}$ in the case of ADAF,
and the main result (i.e., the ``U"-shape curve in the
$\dot{M}-\tilde{r}$ plane as the solution of inner disk size) is
still kept for both $Y_{e}=1/9$ and $Y_{e}=1$. Moreover, although
larger $Y_{e}$ leads to slightly lower value of density, temperature
and pressure for ADAF, the physical properties of the ADAF disk with
low accretion rate beyond equation (\ref{b11}) is close to each
other for the cases of $Y_{e}=1/9$ and 1 (ZD08, Fig. 8).
Furthermore, if we adopt the equilibrium assumption in the ADAF
case, the value of $Y_{e}$ will actually not deviate dramatically
from $0.5$ (i.e., ZD08, the right panel of Fig. 7). Therefore, we
always take $\beta$-equilibrium as an approximation in our
calculation.

Figure 2 shows the structure of the disk for a chosen accretion rate
$0.04 M_{\odot}s^{-1}$ as a function of radius for three values of
$\alpha$. The disk with lower $\alpha$ is denser and thinner with
higher pressure and larger adiabatic index, and has a brighter
neutrino luminosity in most part of the disk except for a part of
the inner disk region, which satisfies the self-similar structure. A
low-$\alpha$ accretion flow with less kinematic viscosity
coefficient $\nu_{k}$ requires a higher surface density $\Sigma$ for
a fixed accretion rate compared to a high-$\alpha$ accretion flow.
We have listed approximate analytic solutions of accretion flows in
various cases in ZD08. We obtain $\rho\propto \alpha^{-1}$,
$P\propto \alpha^{-1}$, $H\propto\alpha^{0}$ for a
radiation-pressure-dominated ADAF,
$\rho\propto(1+Y_{e})^{-12/11}\alpha^{-8/11}$,
$P\propto(1+Y_{e})^{-4/11}\alpha^{-10/11}$, $H\propto
(1+Y_{e})^{4/11}\alpha^{-1/11}$ for a gas-pressure-dominated ADAF,
$\rho\propto(1+Y_{e})^{-9/5}\alpha^{-13/10}$,
$P\propto(1+Y_{e})^{-3/5}\alpha^{-11/10}$, $H\propto
(1+Y_{e})^{3/5}\alpha^{1/10}$ for gas-pressure-dominated NDAF. The
density and pressure always increase with decreasing $\alpha$. These
results are consistent with those shown in Figure 2. The disk region
where is radiation-pressure-dominated is extremely small for low
$\alpha$ (eqs. [22] to [25] in ZD08). Also, as for a low-$\alpha$
disk, the electron fraction $Y_{e}$ is also low, so the disk is
thinner compared to the high-$\alpha$ disk for
gas-pressure-dominated ADAF and NDAF, although the viscosity
parameter contributes an increasing factor $\alpha^{-1/11}$ for the
low-$\alpha$ disk with gas-dominated ADAF.

\subsection{Advection-Dominated Inner Disks}
The entropy-conservation self-similar structure has been used by
Medvedev \& Narayan (2001) and ZD08 to discuss the global accretion
disk structure. However, such a structure is not the only structure
for the neutron-star inner disk, as the entropy-conservation
condition $Q^{+}=Q^{-}$ can be satisfied only for
$\varepsilon\simeq1$ in equation (\ref{a09}). In the case where
$\varepsilon<1$, i.e., the inner disk can only partly release the
heating energy generated by itself and advected from the outer
region, a part of the heating energy in the disk should be still
advected onto the neutron star surface and released from the
surface, and thus the inner disk cannot satisfy the
entropy-conservation self-similar structure. In this case, a part of
the heating energy is still advected into the inner region until it
is released around the neutron star surface. We can approximately
take $Q^{-}=\varepsilon Q^{+}$ in the inner disk for
$\varepsilon\lesssim 1$, and thus the structure of the inner disk
can be described by the ADAF self-similar structure (Spruit et al.
1987, Narayan \& Yi 1994):
\begin{equation}
\rho \propto r^{-3/2},\,\, P \propto r^{-5/2},\,\, v_r \propto
r^{-1/2}.\label{d01}
\end{equation}

In Figure 3, we show the inner disk size for four values of the
energy parameter $\varepsilon$=0.9, 0.7, 0.5 and 0.2. We still fix
$\alpha$=0.1 in this section in order to see the effects of
advection in the inner disk with $\varepsilon<1$ independently. For
the similar reason as in \S 3.1, we also assume the
$\beta$-equilibrium for both the case of ADAF and NDAF disks. We
find that the size of the inner disk becomes smaller for lower
$\varepsilon$. This is because more heating energy can be advected
onto the neutron star surface, and the inner disk size is small
enough to keep energy balance between heating and cooling in the
disk.

Compared with the entropy-conservation self-similar structure, the
size of the advection-dominated inner disk is much larger for a low
accretion rate when most part of the disk is advection-dominated.
When the accretion rate is low, the adiabatic index of the accreting
matter is $\gamma\simeq 4/3$ and the entropy-conservation
self-similar structure (\ref{c01}) can be approximately taken as
$\rho\propto r^{-3}$ and $P\propto r^{-4}$, which requires a more
dramatic change of density and pressure than those of the
advection-dominated inner disk $\rho\propto r^{-3/2}$ and $P\propto
r^{-5/2}$. This difference in structure between entropy-conservation
and advection-dominated inner disks makes the size of the inner
disks be different with each other.

Finally, what we should point out is that the structure of the
advection-dominated self-similar inner disk even with
$\varepsilon\rightarrow 1$ is different from the
entropy-conservation disk with $\varepsilon=1$, since these two
types of self-similar structure are based on different sets of
conservation equations. The advection-dominated structure is based
on the mass continuity, radial momentum and angular momentum
equations, while we do not consider the local energy equation in
which $Q^{+}=Q^{+}_{\rm vis}+Q^{+}_{\rm adv}$ with $Q^{+}_{\rm adv}$
to be difficult to determine locally. We only consider the global
energy equation (\ref{a09}) to calculate the size and structure of
the inner disk. On the other hand, under the energy-conservation
condition $Tds=0$, we can establish the relation $P\propto
\rho^{\gamma}$ from the local energy equation and obtain the
self-similar structure (\ref{c01}) with a combination of the mass
continuity and the radial momentum and local energy equations
(ZD08). However, neither the relation $P\propto \rho^{\gamma}$ nor
the integrated angular momentum equation (\ref{a04}) can be
satisfied in the entropy-conservation solution. In other words, the
angular momentum transfer in the inner disk with the structure
$P\propto \rho^{\gamma}$ cannot be merely due to the viscosity. We
should consider the external torque acted on the disk or the angular
momentum redistribution in the inner disk. We will further discuss
the entropy-conservation in \S5.

\subsection{Inner Disks with Outflows}
In \S3.1 and \S3.2, we do not consider outflows, which may have
important effects on the structure and energy flux distribution of
the entire disk in some cases. Following Narayan \& Yi (1994, 1995)
and Medvedev (2004), if the adiabatic index $\gamma<3/2$, then the
Bernoulli constant of the accretion flow is positive and a
thermally-driven wind or outflow can be produced from the disk.
Therefore, the outflow component can be important in ADAFs. On the
other hand, since the accretion rate of the neutron-star disk is
very large, it is reasonable to assume that the neutron star, which
has a solid surface and is different from the black hole, cannot
accumulate all of the accreting matter at once, and thus an outflow
could be produced near the neutron star surface and exist in the
inner region of the disk.

In this section, we consider the disk structure and neutrino
emission in the disk with a thermally-driven outflow. We consider
two models depending on two mechanisms. In the first model
(hereafter model O1), an outflow is mainly produced in the process
of disk matter accreting onto the surface of a neutron star. We
assume that only the inner disk produces an outflow and the
accretion rate of the outer disk can still be considered as a
constant. We take
\begin{equation}
\dot{M}(r)=\dot{M}_{0}(r/\tilde{r})^{s}\label{e01}
\end{equation}
for the inner disk with $s$ to be the outflow index and
$\dot{M}_{0}$ to be the constant accretion rate in the outer disk.
Second, if an outflow is produced by the accretion process in the
disk, then we consider that the outflow is produced in the entire
disk (hereafter model O2 in this paper), i.e.,
\begin{equation}
\dot{M}(r)=\dot{M}_{0}(r/r_{out})^{s}.\label{e011}
\end{equation}
Strictly speaking, a thermally-driven outflow from the entire disk
is expected in the advection-dominated disk but not in the
neutrino-dominated disk. The winds from the disks which emit a
sufficient high neutrino luminosity are considered to be driven due
to neutrino irradiation (Metzger et al. 2008b). We adopt equation
(\ref{e011}) for all the disks with a wide range of the accretion
rate in model O2 for simplicity, and we also take the index $s$ of
the outflow as a constant. The angular momentum equation with an
outflow can be written as
\begin{equation}
\Sigma
\nu=\frac{1+2s\zeta}{1+2s}\frac{\dot{M}}{3\pi}\left(1-\sqrt{\frac{r_{*}}{r}}\right),\label{e02}
\end{equation}
where $\zeta$ describes a difference between outflow velocity
$v_{outflow,\phi}$ and accretion-disk velocity $v_{\phi}$:
$v_{\phi}-v_{outflow,\phi}=\zeta v_{\phi}$. From equation
(\ref{e02}), we see that if $s=0$ or $\zeta=1$, i.e., there is no
outflow or the toroidal velocity of the outflow is zero and no
angular momentum is taken away by the outflow, then the angular
momentum equation (\ref{e02}) switches back to the common case of
equation (\ref{a04}). In this section, we take $\zeta=0$, i.e.,
$v_{outflow,\phi}\approx v_{acc,\phi}$\footnote{Here we do not
consider the effects of magnetic fields. In fact, the differences in
azimuthal velocity and angular momentum between the outflow and
accretion inflow could also exist when the accretion flow is
governed by a magnetic field ($B_{\phi},B_{z}$) without strong
poloidal component $B_{r}$ (Xie \& Yuan 2008; Bu et al. 2009). The
poloidal magnetic field, however, would cause the outflow to
co-rotate with the disk out to the Alfv\'{e}n radius above the disk
surface. In this section, we take the outflow to co-rotate with the
accretion inflow $v_{outflow,\phi}\approx v_{acc,\phi}$.}.

With the outflow in the inner disk, the self-similar structure
becomes
\begin{equation}
\rho \propto r^{s-3/2},\,\, P \propto r^{s-5/2},\,\, v_r \propto
r^{-1/2}.\label{e03}
\end{equation}
The energy-conservation equation of the inner disk can be rewritten
as
\begin{equation}
\int_{r_{*}}^{\tilde{r}}Q_{\nu}^{-}2\pi
rdr=\varepsilon\left[\int_{r_{*}}^{\tilde{r}}\frac{9}{8}\nu\Sigma\frac{GM}{r^{3}}2\pi
rdr+(1-\bar{f}_\nu)\int_{\tilde{r}}^{r_{out}}\frac{9}{8}\nu\Sigma\frac{GM}{r^{3}}2\pi
rdr\right],\label{e04}
\end{equation}
where we still keep the advection parameter $\varepsilon$. For model
O1 that the outflows merely exist in the inner disk due to the
neutron star surface, using equations (\ref{e01}) and (\ref{e02}),
we derive the energy conservation equation (\ref{e04}) as
\begin{eqnarray}
\frac{1}{\varepsilon}\left(\int_{r_{*}}^{\tilde{r}}Q_{\nu}^{-}2\pi
rdr\right)& = &
\frac{3}{4}\left(\frac{1+2s\zeta}{1+2s}\right)\frac{GM\dot{M}_{0}}{\tilde{r}^{s}}\nonumber
\\ & & \times\left[\frac{1}{1-s}\left(\frac{1}{r_{*}^{1-s}}-\frac{1}{\tilde{r}^{1-s}}\right)
-\frac{r_{*}^{1/2}}{3/2-s}\left(\frac{1}{r_{*}^{3/2-s}}-\frac{1}{\tilde{r}^{3/2-s}}\right)\right]\nonumber
\\ & & +(1-\bar{f}_\nu)\frac{3GM\dot{M}_{0}}{4\tilde{r}}\left[1-\frac{2}{3}
\left(\frac{r_{*}}{\tilde{r}}\right)^{1/2}\right]\label{e05}
\end{eqnarray}
for $s<1$, and
\begin{eqnarray}
\frac{1}{\varepsilon}\left(\int_{r_{*}}^{\tilde{r}}Q_{\nu}^{-}2\pi
rdr\right)& = &
\frac{3}{4}\left(\frac{1+2s\zeta}{1+2s}\right)\frac{GM\dot{M}_{0}}{\tilde{r}}\left\{
\textrm{ln}\left(\frac{\tilde{r}}{r_{*}}\right)
-2\left[1-\left(\frac{r_{*}}{\tilde{r}}\right)^{1/2}\right]\right\}\nonumber
\\ & & +(1-\bar{f}_\nu)\frac{3GM\dot{M}_{0}}{4\tilde{r}}\left[1-\frac{2}{3}
\left(\frac{r_{*}}{\tilde{r}}\right)^{1/2}\right]\label{e06}
\end{eqnarray}
for $s=1$. Here we always consider the outflow index $s\leq1$.

If the outflows exist in the entire disk (model O2), then we have
\begin{eqnarray}
\frac{1}{\varepsilon}\left(\int_{r_{*}}^{\tilde{r}}Q_{\nu}^{-}2\pi
rdr\right)& = &
\frac{3}{4}\left(\frac{1+2s\zeta}{1+2s}\right)\frac{GM\dot{M}_{0}}{r_{out}^{s}}\nonumber
\\ & & \times\left[\frac{1}{1-s}\left(\frac{1}{r_{*}^{1-s}}-\frac{\bar{f}_\nu}{\tilde{r}^{1-s}}\right)
-\frac{r_{*}^{1/2}}{3/2-s}\left(\frac{1}{r_{*}^{3/2-s}}-\frac{\bar{f}_\nu}{\tilde{r}^{3/2-s}}\right)\right]
\label{e07}
\end{eqnarray}
for $s<1$ and
\begin{eqnarray}
\frac{1}{\varepsilon}\left(\int_{r_{*}}^{\tilde{r}}Q_{\nu}^{-}2\pi
rdr\right)& = &
\frac{3}{4}\left(\frac{1+2s\zeta}{1+2s}\right)\frac{GM\dot{M}_{0}}{r_{out}}\nonumber
\\&&\times\left\{\textrm{ln}\left(\frac{r_{out}}{r_{*}}\right)-\bar{f}_{\nu}
\textrm{ln}\left(\frac{r_{out}}{\tilde{r}}\right)
+2\left[\bar{f}_{\nu}+\left(\frac{r_{*}}{\tilde{r}}\right)^{1/2}-2\right]\right\}
\label{e08}
\end{eqnarray}
for $s=1$. We should point out that, since we derive the energy
equations (\ref{e05})-(\ref{e08}) in the disk with outflow using the
angular momentum equation (\ref{e02}) rather than the
entropy-conservation  expression (\ref{c02}), we still use the
self-similar structure (\ref{e03}) to calculate the properties of
the inner disk and the entire disk in the case of $\varepsilon\simeq
1$ both in models O1 and O2.

The top two panels in Figure 4 show the size of the inner disk with
different values of the advection parameter $\varepsilon$. In model
O1, the inner disk is larger for a stronger outflow (larger $s$)
with low accretion rate ($<0.2M_{\odot}s^{-1}$); but for a high
accretion rate ($>0.2M_{\odot}s^{-1}$), the size of the inner disk
decreases with increasing the outflow index $s$. In fact, if the
accretion rate is low, the flow of the outer disk is mainly ADAF and
$\bar{f}_{\nu}\sim 0$. From equations (\ref{e05}) and (\ref{e06}),
most of the energy generated in the outer disk is advected into the
inner region, and the inner disk size is mainly determined by the
self-similar structure (\ref{e03}), which requires a less dramatic
change of density and pressure as functions of radius for higher $s$
or stronger outflow. As a result, the size of the inner disk becomes
larger for a stronger outflow. On the other hand, if the accretion
rate is high, we have $\bar{f}_{\nu}\sim 1$ in the outer disk, the
inner disk size is mainly determined by the heating energy generated
in the inner disk, i.e., the first terms of the right-hand side in
equations (\ref{e05}) and (\ref{e06}), and a stronger outflow
carries away more energy from the disk and allows a smaller size of
the inner disk when the accretion rate is sufficiently high.

The bottom two panels in Figure 4 show the inner disk size in model
O2 in which an outflow exists in the entire disk. The inner disk
structure cannot exist for a high accretion rate
($>0.4M_{\odot}s^{-1}$) except for the weak outflow case ($s=0.2$)
where the inner disk can exist in some range of a high accretion
rate. The change of the inner disk size is significant for a large
outflow index $s$, and even the entire accretion disk can satisfy
the self-similar structure (\ref{e02}) for a sufficiently low
accretion rate. For a high accretion rate ($>0.5M_{\odot}s^{-1}$),
the outflow can take away enough heating energy from the disk, and a
balance between heating and cooling in the entire disk can be built
even without an inner disk. In this range of a high accretion rate,
the structure of the neutron-star disk in model O2 is very similar
to the black hole disk with an outflow (e.g., Kohri et al. 2005).

Figure 5 shows the structure of the disk with different values of
the outflow index $s$=0.2, 0.6 and 1 in model O2. We choose the
accretion rate $\dot{M}=0.2M_{\odot}$ s$^{-1}$ and the energy
parameter $\varepsilon$=0.8. The density and pressure in the disk
decrease with increasing the outflow strength in the entire disk,
and the disk becomes thicker for a stronger outflow. The change of
the temperature is not as obvious as that of the density and
pressure. Most part of the outer region of the disk will be cooler
for a stronger outflow, but the inner disk region can be hotter in
order to release the heating energy advected from the outer region
of the disk.

We make a brief summary in the end of this section. We propose the
inner disk to satisfy the self-similar structure. Table 2 shows the
main results in this section. The inner disk satisfies the
entropy-conservation self-similar structure as in ZD08 for the
energy parameter $\varepsilon\simeq1$, while it becomes an
advection-dominated self-similar region for $\varepsilon<1$. In
outflow model O1, an outflow is produced in the inner disk due to
the prevention effect of the neutron star surface, and model O2
suggests that an outflow exists in the entire disk as in Kohri et
al. (2005). We discuss the size of the inner disk depending on
different structures of the inner disk and outflow and different
values of the viscosity parameter.

\section{Neutrino Annihilation}
\subsection{Calculation Method and Surface Boundary Condition}
Hyperaccreting black hole disks can convert some fraction of the net
accretion energy into the energy of a relativistic outflow or wind
by two general mechanisms: neutrino annihilation and
magnetohydrodynamical (MHD) effects such as the Blandford-Znajek
mechanism or magnetic instabilities. However, for hyperaccretion
disks surrounding neutron stars, the energy conversion mechanism is
mainly due to the neutrino annihilation for magnetic fields at the
neutron star surface $\leq 10^{15}-10^{16}$ G. We consider the
process of pair annihilation $\nu_{i}+\bar{\nu}_{i}\rightarrow
e^{+}+e^{-}$ as the most important interaction for energy
production. As the neutron-star disk is denser, hotter with higher
pressure, and has a brighter neutrino luminosity compared with the
black-hole disk in most cases, the neutrino annihilation efficiency
of the neutron-star disk should be higher than that of the
black-hole disk. Moreover, the surface boundary of the neutron star,
which carries away gravitational-binding energy by neutrino
emission, also makes the neutrino annihilation luminosity of the
neutron-star disk be higher than that of the black-hole counterpart.

In this section, we follow the approximate method used by Ruffert et
al. (1997, 1998) and Popham et al. (1999) to calculate the neutrino
annihilation luminosity, i.e., the vertically-integrated disk is
modeled as a grid of cells in two dimensions ($r, \phi$). The
symbols $\epsilon_{\nu_{i}}^{k}$ and $l_{\nu_{i}}^{k}$ are the mean
neutrino energy and neutrino radiation luminosity with three
different types of neutrino ($i=e,\tau,\mu$) in the cell $k$, and
$d_{k}$ is the distance from the cell $k$ to a certain spatial
point. The neutrino annihilation at any point above the disk is
\begin{eqnarray}
l_{\nu\bar{\nu}}=\sum_{i=e,\mu,\tau}A_{1,i}\sum_{k}\frac{l_{\nu_{i}}^{k}}{d_{k}^{2}}\sum_{k'}\frac{l_{\bar{\nu}_{i}}^{k'}}{d_{k'}^{2}}
(\epsilon_{\nu_{i}}^{k}+\epsilon_{\bar{\nu}_{i}}^{k'})(1-\cos\theta_{kk'})^{2}\nonumber
\\+\sum_{i=e,\mu,\tau}A_{2,i}\sum_{k}\frac{l_{\nu_{i}}^{k}}{d_{k}^{2}}\sum_{k'}\frac{l_{\bar{\nu}_{i}}^{k'}}{d_{k'}^{2}}
\frac{\epsilon_{\nu_{i}}^{k}+\epsilon_{\bar{\nu}_{i}}^{k'}}{\epsilon_{\nu_{i}}^{k}\epsilon_{\bar{\nu}_{i}}^{k'}}(1-\cos\theta_{kk'}),\label{g01}
\end{eqnarray}
where the values of the neutrino cross section constants $A_{1,i}$
and $A_{2,i}$ can be seen in Popham et al. (1999). The total
neutrino annihilation luminosity above the disk can be integrated as
\begin{equation}
L_{\nu\bar{\nu}}=2\pi\int_{r_{*}}^{\infty}dr\int_{H}^{\infty}l_{\nu\bar{\nu}}rdz.\label{g02}
\end{equation}

For a neutron-star disk, we should consider both its different
structure compared with the black-hole disk and the boundary
condition of the neutron star surface layer. The neutrino
annihilation luminosity is not only contributed by neutrinos emitted
from the disk but also from the neutron star surface layer. The
luminosity available to be radiated by the boundary layer at neutron
star surface is (Frank et al. 2002)
\begin{equation}
L_{s}=\frac{1}{4}\dot{M}r_{*}^{2}(\Omega^{2}-\Omega_{*}^{2})-G_{*}r_{*}\simeq
\frac{GM\dot{M}}{4r_{*}}\left(1-\frac{\Omega_{*}}{\Omega}\right)^{2},\label{g03}
\end{equation}
where $\Omega$ and $\Omega_{*}$ are the angular velocity of disk
inner boundary and neutron star surface respectively, and
$G_{*}\simeq \frac{1}{2}\dot{M}r_{*}^{2}(\Omega-\Omega_{*})$ is the
viscous torque acting on the accretion disk. Here we only study the
vertically integrated disk over a half-thickness (height) $H$, and
take $\Omega>\Omega_{*}$. As a result, the luminosity is a function
of accretion rate $\dot{M}$ and neutron star surface angular
velocity $\Omega_{*}$. In the case where the inner disk satisfies
the entropy-conservation self-similar structure, we introduce the
efficiency factor $\eta_{s}$ to measure the energy emitting from the
neutron star surface and rewrite equation (\ref{g03}) as
\begin{equation}
L_{s}=\eta_{s}\frac{GM\dot{M}}{4}\left(\frac{1}{r}-\frac{1}{r_{out}}\right).\label{g02}
\end{equation}
If $\Omega_{*}\sim \Omega$, we have $\eta_{s}\sim 0$ and there is no
emission from the surface layer; or if $\Omega_{*}\sim 0$, we have
$\eta_{s}\sim 1$, which satisfies the Virial condition. We consider
the energy released from the surface layer is mainly carried away by
neutrino emission, and thus the neutrino emission rate and the
temperature at the layer are related by
\begin{equation}
Q_{\nu,s}=\frac{\eta_{s}}{2\pi
r_{*}H_{*}}\frac{GM\dot{M}}{4}\left(\frac{1}{r_{*}}-\frac{1}{r_{out}}\right)\sim
\frac{7}{8}\sigma_{B}T^{4}.\label{g04}
\end{equation}

In the case where the self-similar inner disk is
advection-dominated, we obtain the gravitational energy released by
neutrino emission as
\begin{equation}
L_{s}=\eta_{s}\frac{GM\dot{M}}{4}\left(\frac{1}{r}-\frac{1}{r_{out}}\right)
+(1-\varepsilon)\frac{GM\dot{M}}{4}\left\{\frac{1}{r_{*}}-\bar{f}_\nu\left[1-\frac{2}{3}
\left(\frac{r_{*}}{\tilde{r}}\right)^{1/2}\right]\right\},\label{g05}
\end{equation}
where the second term in the right side of equation (\ref{g05}) is
the heating energy advected from the inner disk to the surface
boundary layer, and we can further write equation (\ref{g05}) as
\begin{equation}
L_{s}=(\eta_{s}+\eta_{s,ADAF})\frac{GM\dot{M}}{4}\left(\frac{1}{r}-\frac{1}{r_{out}}\right),\label{g06}
\end{equation}
where we take the equivalent factor $\eta_{s,ADAF}$ to measure the
energy advected from the advection-dominated inner disk to the
surface boundary. Table 3 shows examples of $\eta_{s,ADAF}$ with
different inner disk structures.

When an outflow exists in the disk, the luminosity at the boundary
surface is dimmer since the accretion rate near the neutron star
surface is lower due to the outflow. We can modify equation
(\ref{g05}) by changing the accretion rate $\dot{M}$ to be
$\dot{M}_{0}(r_{*}/\tilde{r})^{s}$ for model O1 in \S3.3 and
$\dot{M}_{0}(r_{*}/r_{out})^{s}$ for model O2.

\subsection{Results of Annihilation Luminosity}
We calculate the neutrino annihilation luminosity $L_{\nu\bar{\nu}}$
and the total neutrino luminosity $L_{\nu}$ emitted from the disk
and neutron star surface. The results of $L_{\nu\bar{\nu}}$ and
$L_{\nu}$ depend on the value of the viscosity parameter $\alpha$,
the detailed structure of the inner disk, the strength of the
outflow, as well as the neutron star surface boundary condition. We
discuss the effects of these various factors in Figure 6 to Figure
9.

Figure 6 shows the total neutrino annihilation luminosity
$L_{\nu\bar{\nu},NS}$ and the emission luminosity $L_{\nu,NS}$ of a
neutron-star disk with different surface boundary layer conditions
($\eta_{s}$=0 and 0.5), and we compare them with the results of a
black-hole disk. In this figure we take the neutron-star inner disk
structure to satisfy the entropy-conservation self-similar structure
(\ref{c02}). The total luminosity and annihilation luminosity of the
neutron-star disk are brighter than those of a black-hole disk with
the same mass and accretion rate. If we study the neutrino
annihilation from the entire disk without surface boundary emission
($\eta_{s}$=0), the difference between $L_{\nu\bar{\nu},NS}$ and
$L_{\nu\bar{\nu},BH}$ is more significant for a low accretion rate
than for a high accretion rate. We have mentioned the reason in ZD08
that a larger inner disk for a neutron-star disk with a low
accretion rate makes the neutrino luminosity much brighter than its
black hole counterpart, and the annihilation rate also becomes
higher for the neutron-star disk. For a high accretion rate
($>$0.5$M_{\odot}$ s$^{-1}$), the effect of neutrino opacity on
$L_{\nu,BH}$ and $L_{\nu\bar{\nu},BH}$ also be less than that on
$L_{\nu,NS}$ and $L_{\nu\bar{\nu},NS}$. On the other hand, neutrino
emission from the neutron star surface boundary layer
($\eta_{s}$=0.5) makes the annihilation luminosity be more than one
order of magnitude higher than that without boundary emission
($\eta_{s}$=0). $L_{\nu\bar{\nu}}$ reaches 10$^{50}$ ergs s$^{-1}$
when $\dot{M}\sim 1M_{\odot}$ s$^{-1}$ for a black-hole disk or
neutron-star disk with $\eta_{s}$=0, but only needs $\dot{M}\sim
0.1M_{\odot}$ s$^{-1}$ for a neutron-star disk with the boundary
condition $\eta_{s}$=0.5. Therefore, a lower-spin neutron star with
hyperaccreting disk around it could have an obviously higher
annihilation efficiency than that of a higher-spin neutron star. We
will discuss the neutrino annihilation luminosity of a neutron-star
disk in more details in \S 5.

In Figure 7, we show the total neutrino annihilation luminosity of a
neutron star disk with different values of the viscosity parameter
$\alpha$=0.1, 0.01, 0.001 and the energy parameter
$\varepsilon$=0.9, 0.5, 0.1. The disk with a moderate viscosity
parameter ($\alpha$=0.01) has the highest annihilation efficiency
and luminosity for a low accretion rate, and the annihilation
luminosity from a high-$\alpha$ disk ($\alpha$=0.1) becomes the
brightest for an accretion rate $\dot{M}\geq 0.05 M_{\odot}$
s$^{-1}$. As discussed in Figure 2, a low-$\alpha$ disk has a
brighter neutrino luminosity $Q_{\nu}$, but it is thinner than a
high-$\alpha$ disk. These two competitive factors lead to the
annihilation results shown in Figure 7a. Figure 7b shows that the
annihilation efficiency increases with increasing $\varepsilon$.
This is because the disk with lower $\varepsilon$ means more heating
energy in the disk to be advected onto the neutron star surface and
increases the neutrino luminosity of the surface layer, and the
value of $\eta_{s,ADAF}$, which plays an important role in
increasing the annihilation efficiency of the entire disk.

Figure 8 shows the total neutrino annihilation luminosity
$L_{\nu\bar{\nu}}$ of the disk with an outflow. We consider the
results of model O1 in which an outflow only exists in the inner
disk due to the neutron star surface and of model O2 in which an
outflow exists in the entire disk. The difference in annihilation
luminosity with different values of the outflow index $s$ but the
same surface boundary condition is not obvious in model O1. However,
$L_{\nu\bar{\nu}}$ becomes much dimmer for a high outflow index $s$
in model O2, which means a strong outflow in the entire disk and
decreases the neutrino annihilation efficiency significantly.

Furthermore, we study the spatial distribution of the neutrino
annihilation luminosity. Figure 9 illustrates the integrated
annihilation luminosity per cm distribution
\begin{equation}
2\pi r\int_{H}^{\infty}l_{\nu\bar{\nu}}dz\label{h01}
\end{equation}
for two accretion rates $\dot{M}$=0.01$M_{\odot}$ s$^{-1}$ and
0.1$M_{\odot}$ s$^{-1}$ with different physical structures of the
disk. We find that the integrated annihilation luminosity drops
dramatically along the disk radius, and a majority of the
annihilation energy is ejected from the cylindric region above the
disk with $r<3\times10^{6}$cm. The difference of the integrated
luminosity per cm between the black-hole disk and neutron-star disk
is significant for a low accretion rate when the inner
entropy-conservation disk is sufficiently large. The luminosity of
the neutron-star disk with an advecting inner disk can be about four
orders of magnitude higher than its black-hole counterpart, while an
outflow makes the annihilation luminosity above the neutron-star
disk be lower than that of the black-hole disk.

Compared with the black-hole disk, the neutron-star disk could
produce a brighter neutrino flux and more powerful annihilation
luminosity. However, the mass-loss rate driven by neutrino-on-baryon
absorption reactions (Qian \& Woosley 1996) along the polar axis of
the neutron-star disk is also significantly higher than its
black-hole counterpart. This raises the problem of whether
simultaneously more powerful annihilation luminosity and higher mass
loss rate could work together to produce a relativistic jet required
for the GRB phenomena. For example, a heavily baryon-loaded wind
from a new-born neutron star within 100 ms prevents any production
of a relativistic jet (Dessart et al. 2009). However, a weaker wind
above the polar region and spherical asymmetry of the outflow at
late times could make production of a relativistic jet above the
stellar pole become possible. We will discuss this issue in more
details in \S 5.3.

\section{Discussions}

\subsection{Size of the Inner Disk}
In \S 3 we study various self-similar structures of the inner disk.
In this section we further discuss the entropy-conservation
structure. The inner disk can be determined by the self-similar
structure (\ref{c02}) and the energy equation (\ref{a09}). However,
the inner disk with the entropy-conservation structure cannot
satisfy the integrated angular momentum equation (\ref{a04}). There
are two explanations. First, we can consider the term
$d\dot{J}_{ext}/dr\neq 0$ in equation (\ref{a03}) for a steady-state
disk since the angular momentum redistributes in the inner disk
before the entire disk becomes steady-state. As mentioned in ZD08,
because the neutron star surface prevents heating energy from being
further advected inward, the inner disk forms to balance the heating
and cooling energy in the entire disk. As a result, energy and
angular momentum could redistribute in the inner disk. Second,
besides this consideration on angular momentum redistribution, we
can also discuss another type of inner disk with the
entropy-conservation structure discussed by Medvedev \& Narayan
(2000), which still satisfies the angular momentum equation
(\ref{a04}) with $v_{r}\ll v_{K}$ and $\Omega\simeq $constant. The
angular momentum transfer and heating energy generation due to
viscosity can be neglected in the inner disk, as $\Omega$ is
approximately equal to a constant, and the heating energy in the
inner disk is merely the energy advected from the outer disk, i.e.,
$Q^{+}=Q^{+}_{\rm adv}$ in the inner disk. As a result, the global
inner-disk energy conservation equation (\ref{a09}) should be
modified as
\begin{equation}
\int_{r_{*}}^{\tilde{r}}Q_{\nu}^{-}2\pi rdr= \frac{3GM\dot{M}}{4}
\left\{{\frac{1}{3
\tilde{r}}-\frac{\bar{f}_\nu}{\tilde{r}}\left[1-\frac{2}{3}
\left(\frac{r_{*}}{\tilde{r}}\right)^{1/2}\right]}\right\},\label{h01}
\end{equation}
for $\Omega\simeq {\rm constant}=\Omega(\tilde{r})$ in the inner
disk. Compared with equation (\ref{a09}), equation (\ref{h01})
requires a smaller value of $\tilde{r}$ with the same neutron star
mass and accretion rate. Figure 10 shows the value of $\tilde{r}$
for $\alpha$=0.1 and 0.01 with $\Omega\simeq$constant and
$\Omega\propto r^{-3/2}$. The constant angular velocity in the inner
disk decreases the inner disk size compared to the case of
$\Omega\propto r^{-3/2}$. However, in order to get a unified
scenario of the entire disk in various cases, we adopt
$\Omega\propto r^{-3/2}$ for all the self-similar structures in \S
3.

\subsection{Annihilation Results and Disk Geometry}
Several previous studies have been performed to calculate the
neutrino annihilation efficiency above the disk around a black hole
with the effects of disk geometry, gravitational bending, rotation
of central black holes and so on (e.g., Ruffert et al. 1997, 1998;
Popham et al. 1999; Asano \& Fukuyama 2000, 2001; Miller et al.
2003; Birkl et al. 2007). The simulations based on general
relativity show the effects of general relativity such as the Kerr
black holes and bending neutrino geodesics in spacetime increase the
total annihilation rate  by a factor of a few. Also, compared with
the spherical and torus neutrinosphere, the disk neutrinosphere with
the same temperature and neutrino luminosity distribution usually
have the highest annihilation efficiency (Birkl et al. 2007). In
this paper, we still adopt the calculation approach on annihilation
based on Ruffert et al. (1997, 1998) and Popham et al. (1999), and
consider the vertically-integrated Newtonian disk. The most
important results of annihilation calculation in our work is that
the neutron-star disk produce more energetic annihilation luminosity
compared to the black-hole counterpart. Moreover, we consider
neutrinos emitted from the stellar surface region, and the neutrino
emission concentrated in this surface region plays a significant
role in increasing the annihilation luminosity to produce
relativistic ejecta formed by $e^{+}e^{-}$ plasma. On the other
hand, we should note that the effect of surface boundary condition
emission could be reduced if the emission region becomes larger than
we consider in this paper due to outflow or the other cooling
mechanisms rather than neutrino cooling at the surface. Some other
works focusing on neutrino annihilation in supernovae discussed
neutrino emission from the entire spherical neutron star surface
(Cooperstein et al. 1986; Goodman et al. 1987; Salmonson \& Wilson
1999). Therefore, a further work should be done to study the effects
of boundary emission and cooling based on more elaborate
considerations on cooling mechanisms and energy transfer at the
boundary around neutron-star disk. However, as the neutron-star disk
and the surface emission increase the annihilation luminosity more
significantly than the general relativity effects, we conclude our
main results would maintain for more elaborate simulations based on
advanced calculations on neutrino annihilation above the disk.

\subsection{Application to GRB Phenomena}
Energy can be deposited in the polar region of black-hole and
neutron-star disks by neutrino annihilation and MHD processes. We
focus on the annihilation process in this paper. In the black-hole
case, the environment along the polar axis (i.e., the rotation axis
of the disk around neutron star) can be baryon-free. For example,
GRB can commence after the initial collapse for $\sim 15$ s in the
collapsar scenario, when the accretion process in the polar region
becomes sufficiently weak to produce a relatively clean environment
with the mass density $\leq10^{6}$ g cm$^{-3}$ (MacFadyen \& Woosley
1999). In the neutron-star case, however, the enormous neutrino
luminosity of a neutron star would drive appreciable mass-loss from
its surface during the first 20 s of its life (Qian \& Woosley
1996). The wind material will feed the polar region, where a large
amount of annihilation energy is deposited. Dessart et al. (2009)
showed that a newly formed neutron star from the neutron star binary
merger will develop a powerful neutrino-driven wind in the polar
funnel in a few milliseconds after its formation. The mass-loss
associated with the neutrino-driven wind is on the order of
$10^{-3}M_{\odot}$ s$^{-1}$, preventing energy outflow from being
accelerated to relativistic speed and producing a GRB. Their
numerical simulations stops at $t\sim100$ ms, while they considered
that the neutron star will collapse to a black hole quickly.
However, a stable neutron star with much longer lifetime $\gg100$ ms
has also been proposed as the GRB central engine. A rapidly rotating
neutron star formed by the neutron star binary merger (Gao \& Fan
2006) or the merger of a white dwarf binary (King et al. 2001) could
produce extended emission (Metzger et al. 2008b) or X-ray flares
(Dai et al. 2006) about 10-100 s afetr the neutron star birth, and
last for tens of seconds. In the collapsar scenario, a newly formed
neutron star or a magnetar could also form after the initial
collapse or associated supernova explosion, and the lifetime of the
neutron star before collapsing to a black hole can reach as long as
several months to yrs (Vietri \& Stella 1998). Therefore, many GRB
models lead to a longtime ($\gg100$ ms) neutron star, which can
produce other phenomena accompanying the GRB prompt emission. On the
other hand, the accretion timescale can last for longer than 10s in
the collapse process of a massive star, or for several seconds after
the merger of a compact star binary with the disk viscosity
parameter $\alpha<0.1$ (Narayan et al. 2001). Thus we can discuss
the steady-state scenario of a neutron star with lifetime $\gg100$
ms surrounded by a hyperaccreting disk.

For a longtime neutron star, the strengthen of a neutrino-driven
winds above the stellar polar region would drop quickly at late
times, as the total mass-loss rate $\dot{M}\sim t^{-5/3}$, and the
neutrino-driven wind become weaker above the polar region than that
from the low latitudes and midplane region of the neutron star. It
is difficult to calculate the spatial distribution of Lorentz factor
of the outflow material above the polar region precisely in our
present work, because we only use the approximate disk geometry to
calculate the steady-state spatial distribution of the neutrino
annihilation efficiency. We do not simulate the dynamical evolution
of the neutrino-driven wind. However, we can estimate the speed of
outflow material using semi-analytic methods as follows. The mass
loss rate for a thermal neutrino-driven wind can be approximately
given by (Qian \& Woosley 1996)
\begin{equation}
\dot{M}_{\rm
wind}\approx1.14\times10^{-10}C^{5/3}L_{\bar{\nu}_{e},51}\epsilon_{\nu_{e},
MeV}^{10/3}(r/10^{6}\textrm{cm})^{5/3}(M/1.4M_{\odot})^{-2}M_{\odot}\,\textrm{s}^{-1},\label{wind01}
\end{equation}
where $10^{51}$$L_{\bar{\nu}_{e},51}$ergs s$^{-1}$ is the luminosity
of the $\bar{\nu}_{e}$ emission,
$\epsilon_{\nu_{e}}$=1MeV$\epsilon_{\nu_{e}, MeV}$ is the mean
neutrino energy of the neutron star surface, and
$C\approx1+0.733(r/10^{6}\textrm{cm})^{-1}(M/1.4M_{\odot})$. Thus
the typical steady state spherical mass-loss rate due to thermal
neutrino absorption reactions for a neutron star with 10km radius is
on the order of a few $10^{-6}-10^{-5}M_{\odot}$ s$^{-1}$, depending
on the initial configuration of the material above the neutron star
surface (i.e., Qian \& Woosley 1996, Tab. 1). The mass-loss rate
depends on the neutrino luminosity and temperature above the neutron
star surface sensitively. When the neutron star with weak magnetic
field $<10^{15}$ G is surrounded by a hyperaccreting disk, the
neutron star surface should have a higher temperature near the star
midplane than its polar region, as mentioned in the last section \S
4. Thus the neutron star should produce a stronger wind above the
midplane region than above its poles. As a result, the mass-loss
rate via neutrino absorption above the star polar region is
estimated as
\begin{equation}
\dot{M}_{\rm polar}=\dot{M}_{\rm
wind}\left(\frac{\Delta\Omega}{2\pi}\right)f_{asy},\label{wind03}
\end{equation}
where $\Delta\Omega=\int \textrm{cos}\varphi d\varphi d\phi$ is the
solid angle of the polar funnel, and $f_{asy}<1$ is used to measure
the degree of spherical asymmetry. Moreover, the bulk Lorentz factor
of the outflows from the neutron star surface is
\begin{equation}
\Gamma\geq\frac{L_{\nu\bar{\nu}}f_{k}}{\dot{M}_{\rm
polar}c^{2}}=\frac{L_{\nu\bar{\nu}}}{\dot{M}_{\rm
wind}c^{2}}\left(\frac{2\pi f_{k}}{\Delta\Omega
f_{asy}}\right),\label{wind04}
\end{equation}
where $f_{k}$ is the fraction of deposited annihilation energy which
provides kinetic energy of the neutron star wind above the polar
region. We can estimate whether the outflow material from the polar
region of the neutron star surface can be accelerated to a
relativistic speed using equation (\ref{wind04}). Here we take the
bulk Lorentz factor $1<\Gamma<10$ as a mildly relativistic speed,
$10\leq\Gamma<100$ as a moderately relativistic speed and
$\Gamma>100$ as an ultrarelativistic speed. If we take the bulk
Lorentz factor $\Gamma$ as a parameter, the solid angle of the polar
region $\Delta\Omega/2\pi\sim 0.1$, and $f_{k}\sim f_{asy}$, then we
can estimate the upper limit of total thermal mass-loss rate from
the neutron star surface with particular boundary layer conditions,
and compare such limits with the actual mass-loss rate of thermal
neutrino-driven winds calculated by equation (\ref{wind01}). In
Figure 11 we explore the possibility of producing a relativistic
jet. We show the maximum allowed strength of total mass-loss for the
wind material above the polar region being accelerated to
$\Gamma=10$ and $\Gamma=100$ with the chosen boundary layer
condition $\eta_{s}=0$ and $\eta_{s}=0.5$ as in \S 4. The neutrino
emission from the neutron star surface layer increases the neutrino
annihilation luminosity and efficiency, while it increases the
stellar surface mass-loss simultaneously. From Figure 11, we find
that moderately relativistic outflows above the neutron star polar
region are possible under some cases, e.g., the wind material can be
accelerated to $10<\Gamma<100$ above the polar region for the disk
accretion rate $\dot{M}\geq0.08M_{\odot}$ s$^{-1}$ with
$\eta_{s}=0.5$. This result can still be kept for other values of
$f_{k}/f_{asy}$ around unit. However, annihilation process can never
produce any winds with bulk Lorentz factor $\Gamma\geq100$, as the
heavily mass-loaded neutron-star wind precludes ultrarelativistic
speed even for very high annihilation efficiency with sufficiently
bright neutrino emission from the innermost disk radius. Zhang et
al. (2003, 2004) found that relativistic jets formed in the
accreting black hole systems can be collimated by their passage
through the stellar envelope; moderately and even mildly
relativistic jets can be partly accelerated to an ultrarelativistic
speed after they break out in the massive star, because the jets'
internal energy can be converted to kinetic energy after jet
breakout. Such jet-stellar-envelope interactions also happen around
the hyperaccreting neutron star systems, although the neutron-star
disk systems can be surrounded by a cavity $\sim 10^{9}$ cm inside
the progenitor stars with the stellar radius of several $10^{10}$ cm
(e.g., Bucciantini et al. 2008, 2009). In the compact star binary
merger scenario, on the other hand, a moderately relativistic jet is
possible to produce a short-duration GRB.

As a result, the neutron-star disk can produce a sufficiently larger
annihilation luminosity than its black-hole counterpart. However, a
neutrino-driven outflow from the newly formed neutron star at early
times ($\sim100$ ms) is so heavily mass-loaded that it can in no way
be accelerated to relativistic speed. For a longtime neutron star,
however, mass-loss becomes weaker above the stellar pole than above
the low latitudes and midplane. Thus a moderately relativistic jet
can be produced in a hyperaccreting neutron-star system with
sufficiently high disk accretion rate and bright boundary emission
(e.g., the accretion rate $\dot{M}\geq0.08M_{\odot}$ s$^{-1}$ for
$\eta_{s}=0.5$). Also, the jet can be collimated by
jet-stellar-envelope interactions, and partly accelerated to an
ultrarelativistic speed after jet breakout if the neutron star forms
through stellar collapse. Therefore, some hyperaccreting
neutron-star system can produce GRBs, which are more energetic than
those from the black-hole systems. We know that some long-duration
GRBs can reach a peak luminosity of $\sim 10^{52}$ ergs s$^{-1}$
(e.g. GRB 990123, Kulkarni et al. 1999), which requires a very high
accretion rate $\sim 10 M_{\odot}$ s$^{-1}$ as well as a much more
massive disk ($\geq 10 M_{\odot}$) around a black hole compared to
the typical disk or torus mass $0.01-1 M_{\odot}$ if the energy is
provided by neutrino annihilation above the black-hole disk.
However, if we consider the neutron-star disk with the surface
boundary emission (e.g. $\eta_{s}\sim 0.5$), we only need an
accretion rate $\sim 1 M_{\odot}$ s$^{-1}$ onto the neutron star,
which is one order of magnitude less than that of the black-hole
disk, and the disk mass can be $\sim 1 M_{\odot}$ for a long burst
with the peak luminosity $10^{52}$ ergs s$^{-1}$ in a time of about
1s. However, other reasons or mechanisms such as the jet effect to
reduce the total burst energy or the magnetic mechanism rather than
neutrino annihilation to provide the GRB energy have been introduced
into the GRB central engine models. Therefore, it is necessary to
search new important observational evidence to show the existence of
a central neutron star rather than black hole surrounded by a
hyperaccreting disk. X-ray flares after GRBs may be a piece of
evidence, which shows that an activity of the central neutron star
after the burst may be due to magnetic instability and reconnection
effects in differentially-rotating pulsars (Dai et al. 2006).
However, more studies should be done to compare other models of
X-ray flares (King et al. 2005; Proga \& Zhang 2006; Perna et al.
2005; Lee \& Ramires-Ruiz 2007, Lazzati et al. 2008) with the
differential-rotating pulsar model and show more effects of such a
magnetized pulsar. For example, the spin-down power of a magnetar
probably explains the peculiar optical to X-ray integrated
luminosity of GRB 060218 (Soderberg et al. 2006).

Furthermore, we propose that other GRB-like events may be produced
by hyperaccreting neutron star systems, if winds fail to reach a
proper relativistic speed. As mentioned by MacFadyen et al. (2001),
a mildly or moderately relativistic jet may lead to X-ray flashes,
which are less energetic than normal GRBs. This will also happen in
the neutron-star case. Another possible result is that a heavily
mass-loaded wind with nonrelativistic speed could produce a bright
SN-like optical transient event (Kohri et al. 2005; Metzger et al.
2008a).

\subsection{GRB-SN Association}
Besides GRBa and GRB-like phenomena, we think a nonrelativistic or
mildly relativistic outflow from a hyperaccreting disk and neutron
star surface may feed a supernova explosion associated with a GRB.
The discovery of connection between some GRBs and supernovae has
inspired studies of the origin of GRB-SN association (e.g., Iwamota
et al. 1998; MacFadyen \& Woosley 1999; Zhang et al. 2004; Nagataki
et al. 2006, 2007; Mazzali et al. 2006). As discussed in Kohri et
al. (2005), an outflow from the hyperaccreting disk is possible to
provide a successful supernova with both prompt explosion or delayed
explosion. Therefore, it is reasonable to propose a general scenario
for the origin of GRB-SN connection: some hyperaccreting disks with
outflows around compact objects are central engines of GRBs or XRFs
companied by supernovae. The thermal-driven outflow energy from the
disk provides a part of or even a majority of the kinetic energy of
a supernova, and neutrino annihilation from the disk provides the
energy of a GRB if the accretion rate is sufficiently high. Now we
use the results in \S3.3 and \S 4.2 to discuss the energy of the
outflow from the disk, and compare it with the neutrino annihilation
luminosity and energy above the disk. The upper limit of the energy
rate carried away by the outflow can be estimated by subtracting the
heating energy rate generated in the disk from the ideal heating
energy rate of the disk without outflow, i.e., the maximum energy
rate of the outflow is
\begin{equation}
\dot{E}_{o, max}\sim \frac{3GM\dot{M}}{4} \left\{{\frac{1}{3
r_{*}}-\frac{1}{r_{out}}\left[1-\frac{2}{3}
\left(\frac{r_{*}}{r_{out}}\right)^{1/2}\right]}\right\}
-\int_{r_{*}}^{r_{out}}\frac{9}{8}\nu\Sigma\frac{GM}{r^{3}}2\pi
rdr,\label{h02}
\end{equation}
and the total actual outflow energy can be considered to be $\sim
0.1-1$ fraction of the maximum outflow energy. In Table 4 we list
various energy rates, including the heating energy rate in the disk,
the maximum energy injection rate to an outflow and the neutrino
annihilation rate above the disk. We choose the accretion rate to be
0.3$M_{\odot}$ s$^{-1}$ in both models O1 and O2 in \S 3.3. Compared
with model O1, the neutron-star disk with an outflow from the entire
disk (model O2) produces higher outflow energy but less neutrino
annihilation rate. If the disk mass around a neutron star is $\sim
1M_{\odot}$, then the maximum outflow energy is $\sim 10^{51}$ ergs
in model O1, and $\sim 10^{52}$ ergs in model O2, but the
annihilation luminosity in model O2 would be one or two orders of
magnitude less than that in model O1. Besides the case of a neutron
star, the black-hole disk with an outflow can provide the same order
of outflow energy ejecta as in model O2, but even dimmer
annihilation luminosity than model O2. Therefore, further studies of
an energy relation between GRBs and supernovae in GRB-SN events
could distinguish between the neutron-star disk models (model O1 and
O2) and the black-hole disk model.

\subsection{Effects of Magnetic Fields}
In this paper, we do not consider the effect of magnetic fields. As
mentioned in ZD08, the high magnetic fields of central neutron stars
or magnetars $> 10^{15}-10^{16}$G could play a significant role in
the global structure of the disk as well as various microphysical
processes in the disk. Moreover, compared to the neutron star
surface, which produces $e^{+}e^{-}$ jets and outflows, the central
magnetars could be considered as a possible source to produce
magnetically-dominated outflows and collimated jets with
ultrarelativistic bulk Lorentz factors (e.g., Usov 1992; Lyutikov
2006; Uzdensky \& MacFadyen 2007; Tchekhovskoy et al. 2008; Metzger
et al. 2008a, Bucciantini et al. 2008, 2009). For example,
Bucciantini et al. (2008, 2009) modeled the interaction between the
wind from a newly formed rapidly rotating protomagnetar and the
surrounding progenitor. The free-flowing wind from protomagnetar is
not possible to achieve simultaneously collimation and acceleration
to high Lorentz factor. However, a bubble of relativistic plasma and
a strong toroidal magnetic field created by the magnetar wind
shocking on the surrounding stellar envelope can work together to
drive a relativistic jet, which is possible to produce a long GRBs
and the associated Type Ic supernova. Metzger et al. (2008a) showed
that protomagnetars are capable to produce neutron-rich long GRB
outflows for submillisecond rotation period $P\leq0.8$ ms. Besides
the study on the central magnetar, the properties of magnetized NDAF
disk has also been studied. Recently Lei et al. (2009) investigated
the properties of the NDAF with the magnetic torque acted between
the central black hole and the disk. The neutrino annihilation
luminosity can be increased by one  order of magnitude higher for
accretion rate $\sim0.5M_{\odot}$\,s$^{-1}$, and the disk becomes
thermally and viscously unstable in the inner region. Therefore, it
is also interesting to study the effects of ultra-highly magnetic
fields of neutron stars or magnetars on hyperaccreting disks.

\section{Conclusions}
In this paper we have studied the structure of a hyperaccreting disk
around a neutron star based on the two-region scenario, and
calculate the neutrino annihilation luminosity above the disks with
different structures. The neutron-star disk model is still
Newtonian, vertically-integrated with one-dimensional variable
radius $r$, and based on the $\alpha$-prescription. The accretion
rate $\dot{M}$ is the basic quantity to determine the properties of
the disk and the annihilation luminosity, and we also discuss the
effects of the energy parameter $\varepsilon$ of the inner disk
(eqs. [13], [35] to [39]), the viscosity parameter $\alpha$, the
outflow structure and strength (eqs. [31], [32]), the neutrino
emission for the stellar surface layer to increase the total
annihilation energy rate (eqs. [42] to [46]).

We adopt the self-similar structure to describe the inner disk, in
which the heating mechanism is different from the outer disk. Table
2 shows the inner disk structure and size in various cases depending
on the value of $\varepsilon$ and the properties of the outflow from
the disk. We introduce two disk outflow models in \S 3.3. In model
O1, we consider the outflow is mainly from the inner disk, while
model O2 suggests that an outflow exists in the entire disk. In \S
5.1, we also discuss the other possibilities of the
entropy-conservation self-similar structure of the inner disk.

Compared to a high-$\alpha$ disk ($\alpha\sim 0.1$), the size of a
low-$\alpha$ disk is smaller for a low accretion rate ($\leq 0.1
M_{\odot}$ s$^{-1}$) and increases dramatically with increasing
accretion rate (Fig. 1). A low-$\alpha$ disk is denser, thinner with
higher pressure and larger adiabatic index, and has a brighter
neutrino luminosity (Fig. 2). The size of the inner disk which
satisfies the advection-dominated self-similar structure for
$\varepsilon<1$ becomes smaller for lower $\varepsilon$, and is much
larger compared to the entropy-conservation inner disk for a low
accretion rate ($\leq 0.1 M_{\odot}$ s$^{-1}$, Fig. 3). In outflow
model O1, the inner disk is larger for a stronger outflow with low
accretion rate, but its size decreases with increasing the outflow
index $s$ for a high accretion rate. Moreover, the inner disk would
not exist in outflow model O2 when the accretion rate $\geq
0.5M_{\odot}$ s$^{-1}$ (Fig. 4). The outflow in the entire disk
decreases the density and pressure, but increases the thickness of
the disk (Fig. 5).

The neutrino annihilation luminosity above the neutron-star disk
$L_{\nu\bar{\nu},NS}$ is higher than the black-hole disk
$L_{\nu\bar{\nu},BH}$, and the difference between
$L_{\nu\bar{\nu},NS}$ and $L_{\nu\bar{\nu},BH}$ is significant for a
low accretion rate due to the different disk structure and neutrino
luminosity between them (Fig. 6). The neutrino emission from the
neutron star surface boundary layer is produced in the process of
disk matter accreting onto the surface, and the boundary emission
can increase the total neutrino annihilation rate above the disk
significantly for about one order of magnitude higher than the disk
without boundary emission (Fig. 6, Fig. 7, Fig. 8). The disk with an
advection-dominated inner disk could produce the highest neutrino
luminosity while the disk with an outflow from the entire disk
(model O2) produces the lowest annihilation luminosity (Fig 7, Fig
8, Fig 9). We show that the annihilation luminosity can reach
$10^{52}$ergs s$^{-1}$ when the accretion rate $\sim 1 M_{\odot}$
s$^{-1}$ for neutron-star disks, while the higher accretion rate
$\sim 10 M_{\odot}$ s$^{-1}$ is needed to reach $10^{52}$ ergs
s$^{-1}$ for black-hole disks. Therefore, the neutron-star disk can
produce a sufficiently large annihilation luminosity than its
black-hole counterpart. Although a heavily mass-loaded outflow from
the neutron star surface at early times of neutron star formation
prevents the outflow material from being accelerated to a high bulk
Lorentz factor, an energetic relativistic jet can be produced above
the stellar polar region at late times if the disk accretion rate
and the neutrino emission luminosity from the surface boundary are
sufficiently high. Such relativistic jet may be further accelerated
by jet-stellar-envelope interaction and produce the GRB or GRB-like
events such as X-ray flashes (XRFs).

The outflow from the advection-dominated disk and low latitudes of
the neutron star surface can be considered to provide the energy and
sufficient $^{56}$Ni for successful supernova explosions associated
with GRBs or XRFs in some previous works. However, the energy
produced via neutrino annihilation above the advection-dominated
black-hole disk is usually not sufficient for relativistic ejecta
and GRBs. On the other hand, the advection-dominated disk around a
neutron star can produce a much higher annihilation luminosity
compared to the black-hole disk. Outflow model O2 produces a higher
neutrino annihilation rate but less outflow energy compared to model
O1, while the black-hole disk could provide the same order of
outflow energy but even less annihilation energy compared to model
O2. Therefore, observations on GRB-SN connection would further
constrain these models between hyperaccreting disks around black
holes and neutron stars with outflows.

\section*{Acknowledgements}
We would like to thank the anonymous referee for his/her very useful
comments that have allowed us to improve our paper significantly.
This work is supported by the National Natural Science Foundation of
China (grants 10221001, 10640420144 and 10873009) and the National
Basic Research Program of China (973 program) No. 2007CB815404.

\newpage
\begin{table}
\begin{center}
Notation and definition of some quantities in this paper
\begin{tabular}{|l|l|r|r|r|r|}
\hline
notation & definition & \S/Eq. \\
\hline
 $\varepsilon$ & energy parameter& \S2.1, eq.(\ref{a09})\\
 $\gamma$ & adiabatic index of the accreting matter& \S2.2, eq.(\ref{b07})\\
 $s$ & outflow index& \S3.3, eq.(\ref{e01})\\
 $\zeta$ & toroidal velocity difference between outflow and accretion disk& \S3.3, eq.(\ref{e02})\\
 $\eta_{s}$ & efficiency factor to measure the surface emission& \S4.1, eq.(\ref{g02})\\
 $\eta_{s,ADAF}$ & efficiency factor to measure the energy advected from the disk& \S4.1, eq.(\ref{g06})\\
\hline
\end{tabular}
\caption{}
\end{center}
\end{table}

\begin{table}
\begin{center}
Self-similar structure and size of the inner disk in various cases
\begin{tabular}{rccccrc}
\hline \hline
Cases & Inner disk self-similar structure & Inner disk size & \\
\hline
Case 1: $\varepsilon\simeq$1, no outflow & entropy-conservation structure (\ref{c02})  & Figure 1\\
Case 2: $\varepsilon<$1, no outflow & advection-dominated structure (\ref{d01})  & Figure 3\\
Case 3: outflow model O1 & outflow structure (\ref{e03})  & Figure 4 top panels\\
Case 4: outflow model O2 & outflow structure (\ref{e03})  & Figure 4 bottom panels\\
\hline \hline
\end{tabular}
\caption{We discuss case 1 in \S 3.1, case 2 in \S 3.2, case 3 and
case 4 in \S 3.3 with different inner disk structures based on
energy and mass transfer in the entire disk.}
\end{center}
\end{table}
\begin{table}
\begin{center}
Equivalent factor $\eta_{s,ADAF}$ with different inner disk
structures
\begin{tabular}{rccccrc}
\hline \hline $\eta_{s,ADAF}$ & 0.03 & 0.1 & 0.3 & 0.5 & 1.0 & 3.0\\
&($M_{\odot}$ s$^{-1}$)\\
\hline
$\varepsilon$=0.9 & 9.07e-2 & 5.25e-2 & 4.22e-2 & 5.56e-2 & 7.75e-2 & 9.56e-2\\
$\varepsilon$=0.5 & 0.419 & 0.196 & 0.159 & 0.149 & 0.180 & 0.446\\
$\varepsilon$=0.2 & 0.675 & 0.496 & 0.219 & 0.178 & 0.218 & 0.382\\
\hline
O1: $s$=0.2 & 0.228 & 0.113 & 7.84e-2 & 8.72e-2 & 0.107 & 0.202\\
$s$=0.6 & -- & 0.215 & 0.115 & 0.113 & 0.119 & 0.209\\
$s$=1.0 & -- & 0.315 & 0.199 & 0.168 & 0.203 & 0.336\\
\hline
O2: $s$=0.2 & 0.238 & 0.104 & 7.95e-3 & -- & -- & --\\
$s$=0.6 & -- & 0.428 & 5.79e-4 & -- & -- & --\\
$s$=1.0 &-- & -- & 0.639 & 4.36e-4 & -- & --\\
\hline \hline
\end{tabular}
\caption{Equations (\ref{g05}) and (\ref{g06}) give the value of
$\eta_{s,ADAF}$. We choose cases of the advection-dominated inner
disk with $\varepsilon$=0.9, 0.5, 0.2, and models O1 and O2 with
$s$=0.2, 0.6 1.0 to calculate $\eta_{s,ADAF}$.}
\end{center}
\end{table}

\begin{table}
\begin{center}
Comparison of energy rates between outflow and neutrino annihilation
\begin{tabular}{rccccrc}
\hline \hline Cases & disk heating rate & max outflow energy&
$L_{\nu\bar{\nu}}$ ($\eta_{s}$=0) & $L_{\nu\bar{\nu}}$
($\eta_{s}$=0.5)& \\& ($10^{51}$ergs s$^{-1}$)& rate ($10^{51}$ergs
s$^{-1}$) & ($10^{51}$ergs s$^{-1}$) & ($10^{51}$ergs s$^{-1}$)\\
\hline
model O1: $s$=0.2 & 25.3 & 1.09 & 8.37e-2 & 2.24\\
$s$=0.6 & 24.7 & 1.72 & 0.104 & 1.74\\
$s$=1.0 & 18.4 & 7.95 & 0.179 & 1.65\\
\hline
model O2: $s$=0.2 & 14.6 & 11.8 & 1.42e-2 & 0.522\\
$s$=0.6 & 6.50 & 19.9 & 4.84e-3 & 2.54e-2\\
$s$=1.0 & 4.07 & 22.3 & 3.48e-3 & 5.99e-3\\
\hline \hline
\end{tabular}
\caption{We consider model O1 and O2 with the outflow index $s$=0.2,
0.6, 1.0, the accretion rate $\dot{M}=0.3M_{\odot}$ s$^{-1}$ and the
energy parameter $\varepsilon$=0.8.}
\end{center}
\end{table}

\newpage
\begin{figure}
\resizebox{\hsize}{!} {\includegraphics{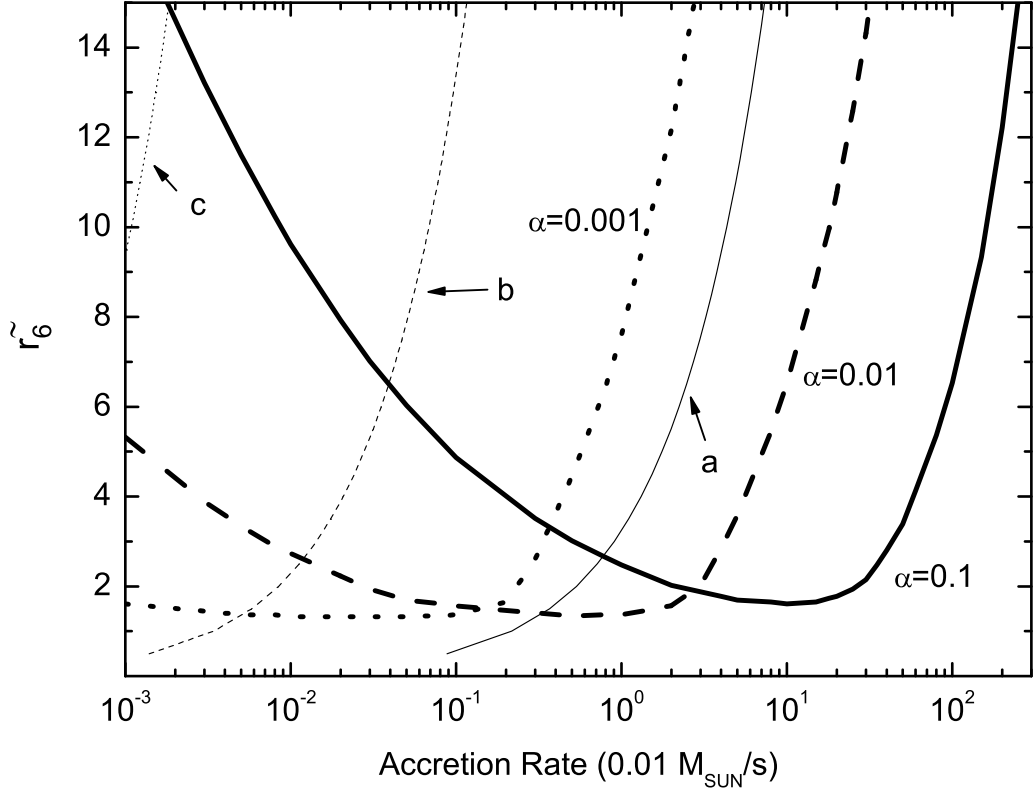}} \caption{The
radius $\tilde{r}$ ($\tilde{r}_{6}$, in units of $10^{6}$ cm) of the
boundary layer between the inner and outer disks as a function of
accretion rate with different values of the viscosity parameter
$\alpha$=0.1 (thick solid line), 0.01 (thick dashed line) and 0.001
(thick dotted line). The three thin lines labeled ``a", ``b", and
``c" are the characteristic curves of $\beta$-equilibrium with three
values of $\alpha$=0.1, 0.01 and 0.001 respectively. Each
characteristic curve divides the $\dot{M}-\tilde{r}$ parameter plane
into two regions with a chosen $\alpha$, and $\beta$-equilibrium can
be established in the right region respectively.}
\end{figure}

\newpage
\begin{figure}
\resizebox{\hsize}{!} {\includegraphics{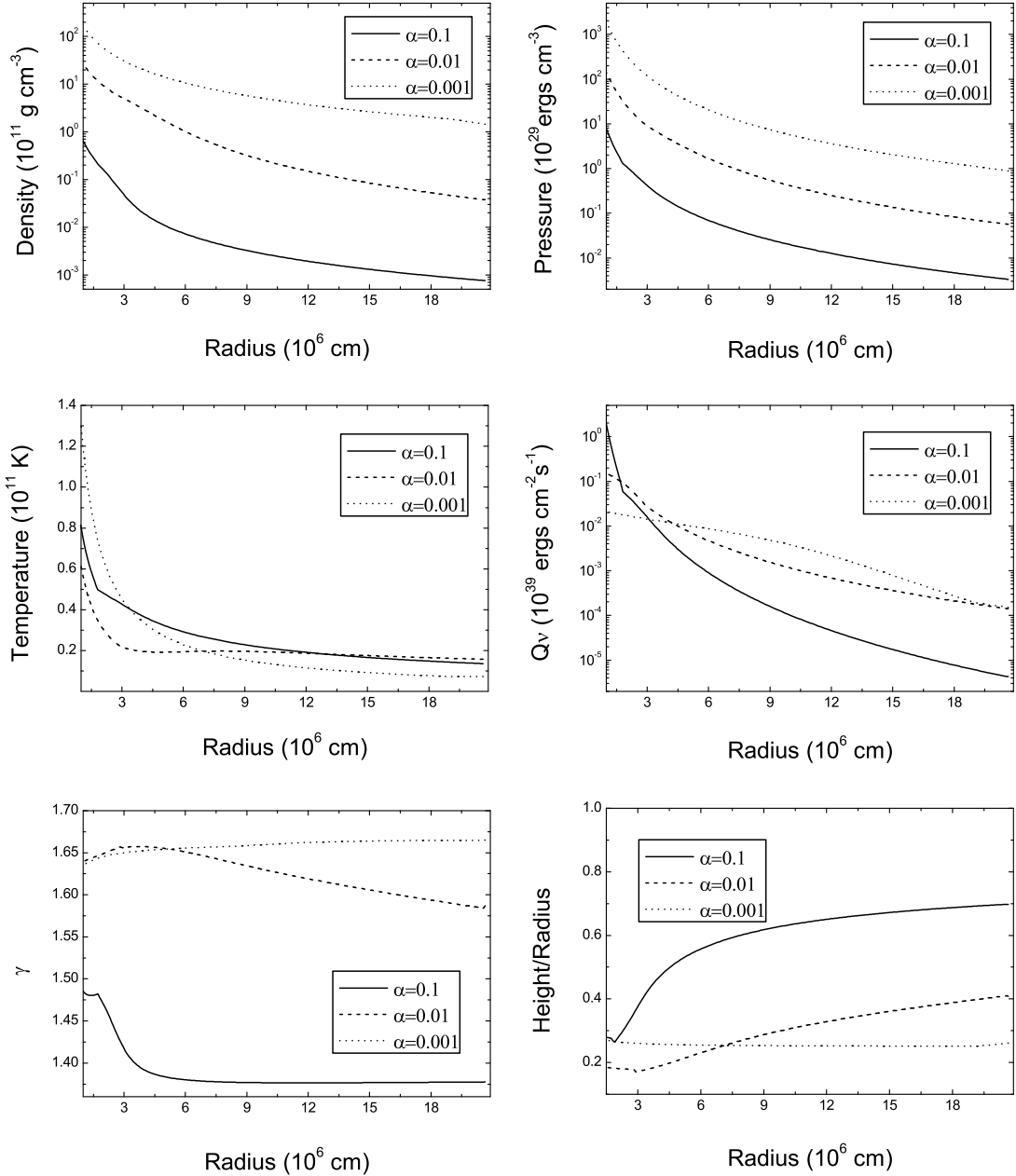}} \caption{The
density, pressure, temperature, neutrino luminosity per unit area,
adiabatic index $\gamma$ and height (half-thickness of the disk) of
the entire disk with different values of the viscosity parameter
$\alpha$=0.1 (solid line), 0.01 (dashed line) and 0.001 (dotted
line) with a fixed accretion rate $\dot{M}=0.04M_{\odot}$ s$^{-1}$.
}
\end{figure}

\newpage
\begin{figure}
\resizebox{\hsize}{!} {\includegraphics{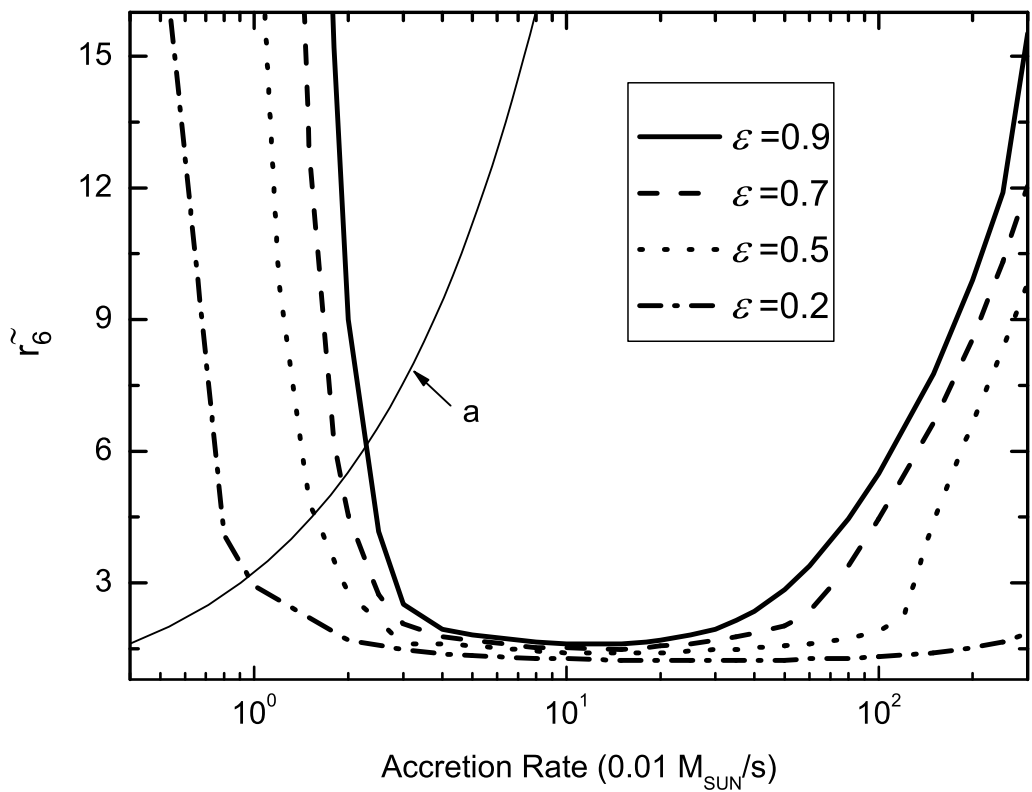}} \caption{The
radius $\tilde{r}$ (in units of $10^{6}$ cm) between the inner and
outer disks as a function of accretion rate with different values of
the energy parameter $\varepsilon$=0.9, 0.7, 0.5 and 0.2. The thin
line labeled ``a" is the characteristic curve of equilibrium as in
Fig. 1 with $\alpha$=0.1.}
\end{figure}

\newpage
\begin{figure}
\resizebox{\hsize}{!} {\includegraphics{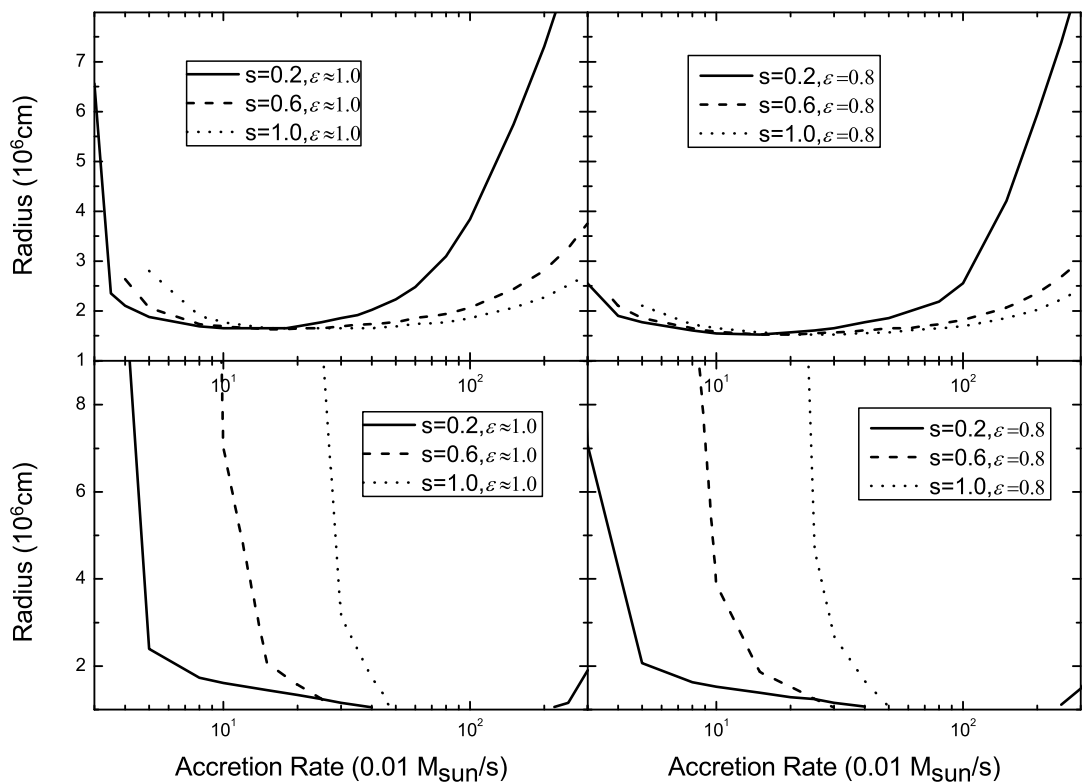}} \caption{(a) {\em
Top panels}: the radius $\tilde{r}$ between the inner and outer disk
in model O1 with different parameter sets of the outflow index
$s$=0.2, 0.6, 1 and the energy parameter $\varepsilon\simeq$ 1, 0.8.
(b) {\em Bottom panels}: the radius $\tilde{r}$ in model O2 with the
same parameter sets ($s, \varepsilon$) as in the top two panels.}
\end{figure}

\newpage
\begin{figure}
\resizebox{\hsize}{!} {\includegraphics{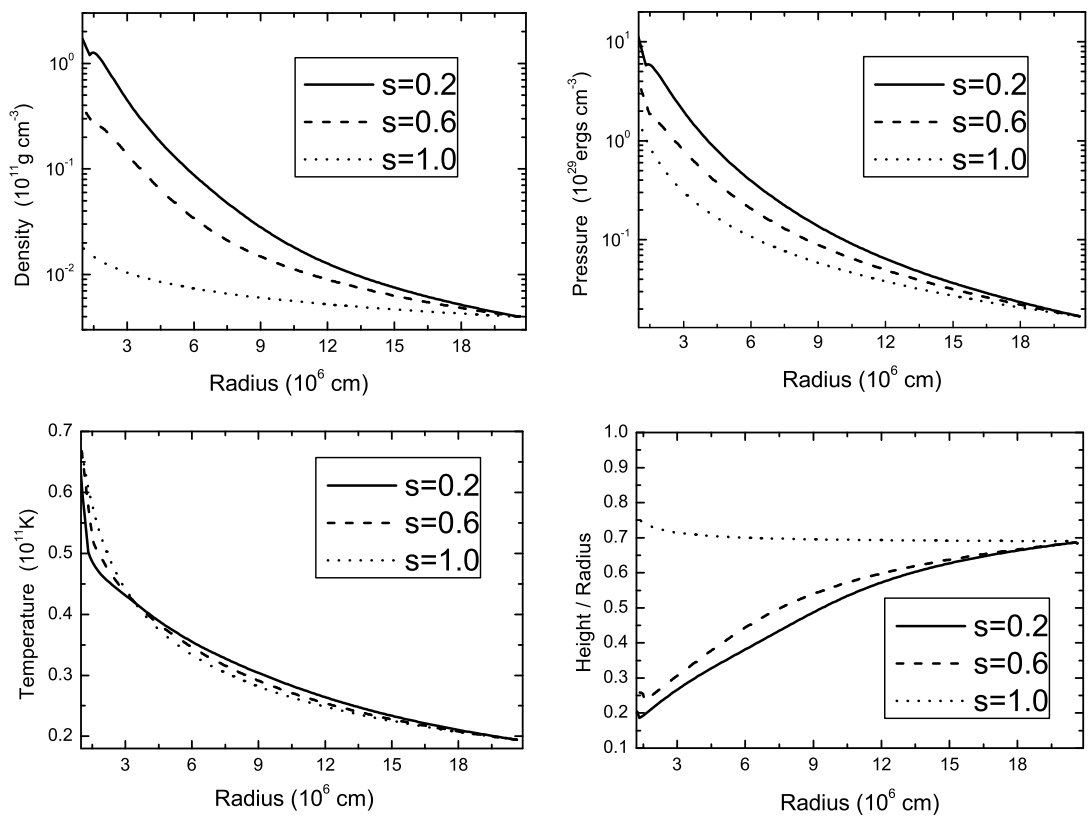}} \caption{Density,
pressure, temperature and height (half-thickness) as functions of
radius for different values of the outflow index $s$=0.2, 0.6, 1,
where we take the accretion rate as 0.2$M_{\odot}$ $\rm s^{-1}$ and
the energy parameter $\varepsilon$=0.8.}
\end{figure}

\newpage
\begin{figure}
\resizebox{\hsize}{!} {\includegraphics{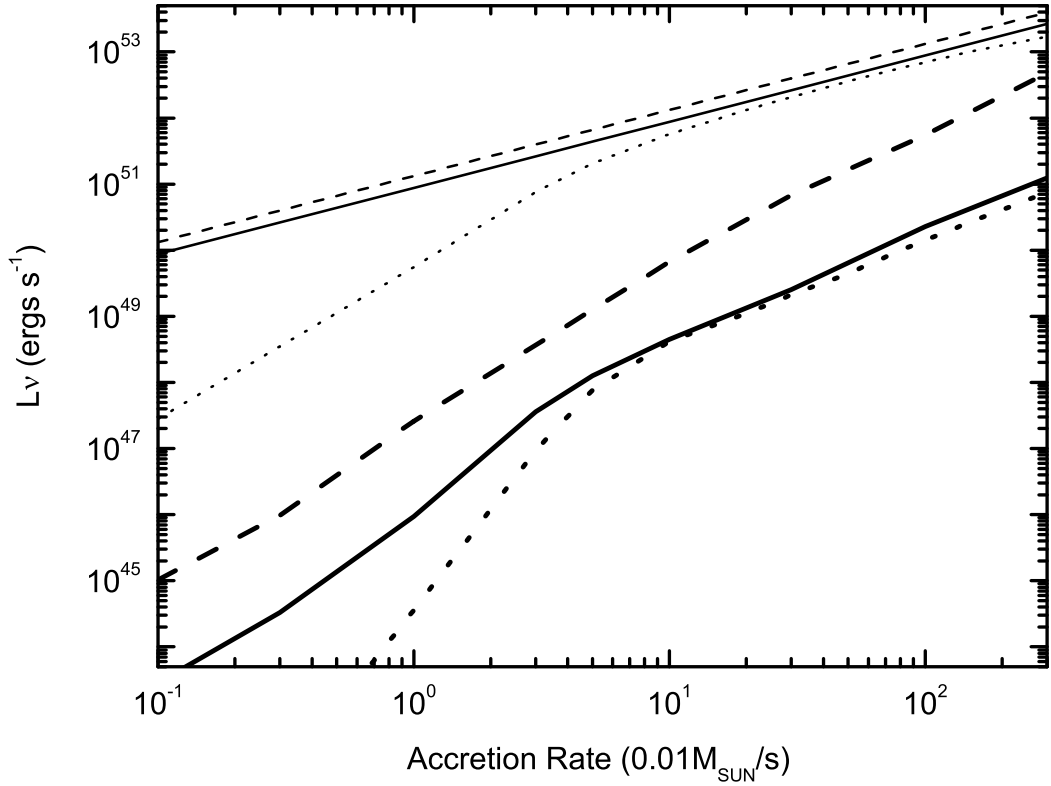}}\caption{Neutrino
annihilation luminosity $L_{\nu\bar{\nu}}$ ({\em thick lines}) and
total neutrino emission luminosity $L_{\nu}$ ({\em thin lines}) as
functions of accretion rate. The solid lines correspond to the
neutron-star disk with the entropy-conservation inner disk and the
boundary layer emission efficiency $\eta_{s}=0$, the dashed lines to
the neutron-star disk with the boundary condition $\eta_{s}=0.5$,
and the dotted lines to the black-hole disk.}
\end{figure}

\newpage
\begin{figure}
\resizebox{\hsize}{!} {\includegraphics{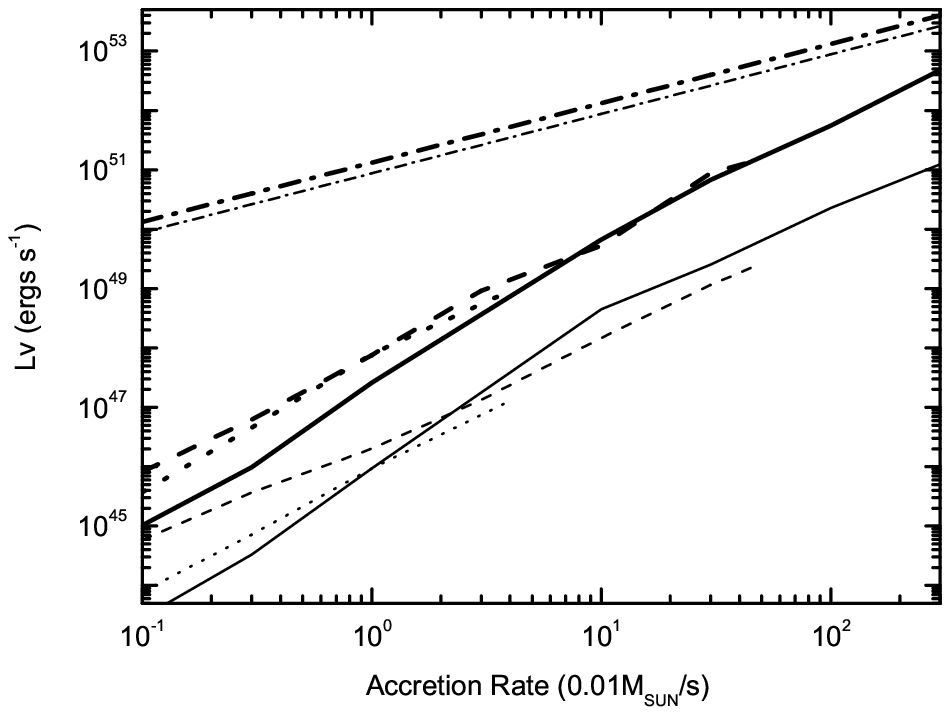}
\includegraphics{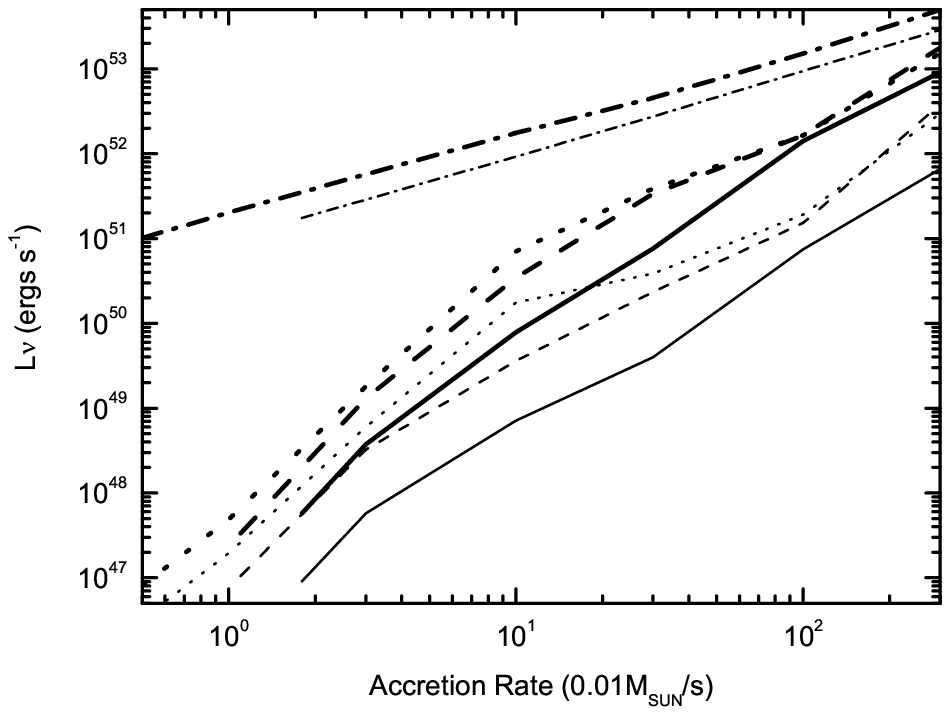}}
\caption{Neutrino annihilation luminosity $L_{\nu\bar{\nu}}$ and the
total neutrino luminosity $L_{\nu}$ with different values of the
viscosity parameter $\alpha$ and energy parameter $\varepsilon$. The
thin lines correspond to $\eta_{s}=0$ and thick lines to
$\eta_{s}=0.5$. (a) {\em Left panel}: $L_{\nu\bar{\nu}}$ with
$\alpha$=0.1 ({\em solid line}), 0.01 ({\em dashed line}), 0.001
({\em dotted line}) and total luminosity $L_{\nu}$ ({\em dash-dotted
line}), where we take $\varepsilon=1$ and the entropy-conservation
inner disk structure. (b) {\em Right panel}: $L_{\nu\bar{\nu}}$ of
the neutron-star disk with advection-dominated inner disk and
$\varepsilon$=0.9 ({\em solid line}), 0.5 ({\em dashed line}), 0.1
({\em dotted line}) and the total luminosity $L_{\nu}$ ({\em
dash-dotted line}).}
\end{figure}

\newpage
\begin{figure}
\resizebox{\hsize}{!} {\includegraphics{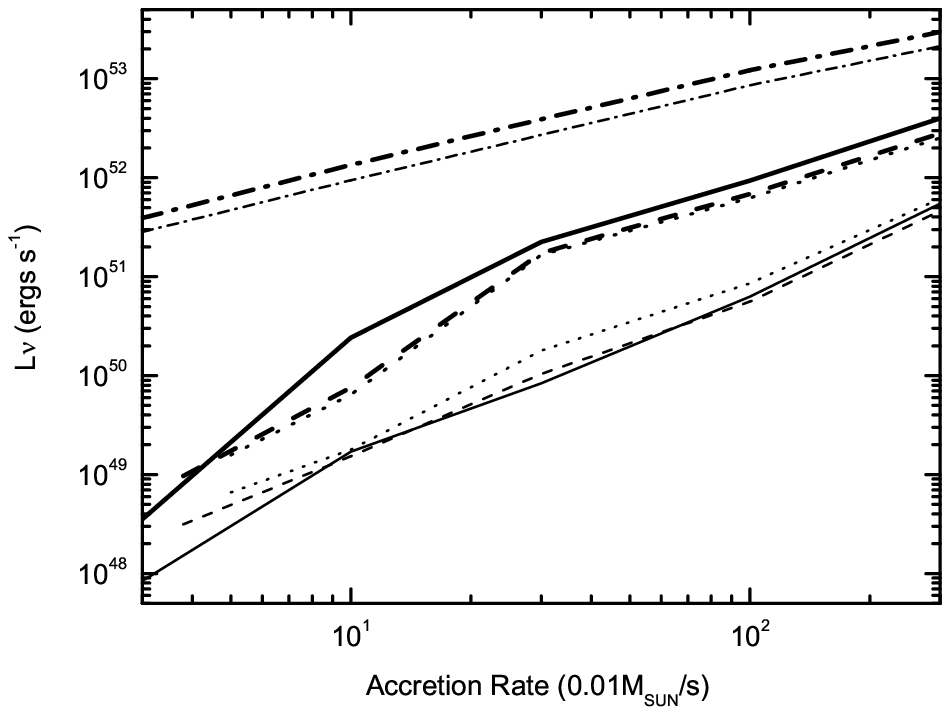}
\includegraphics{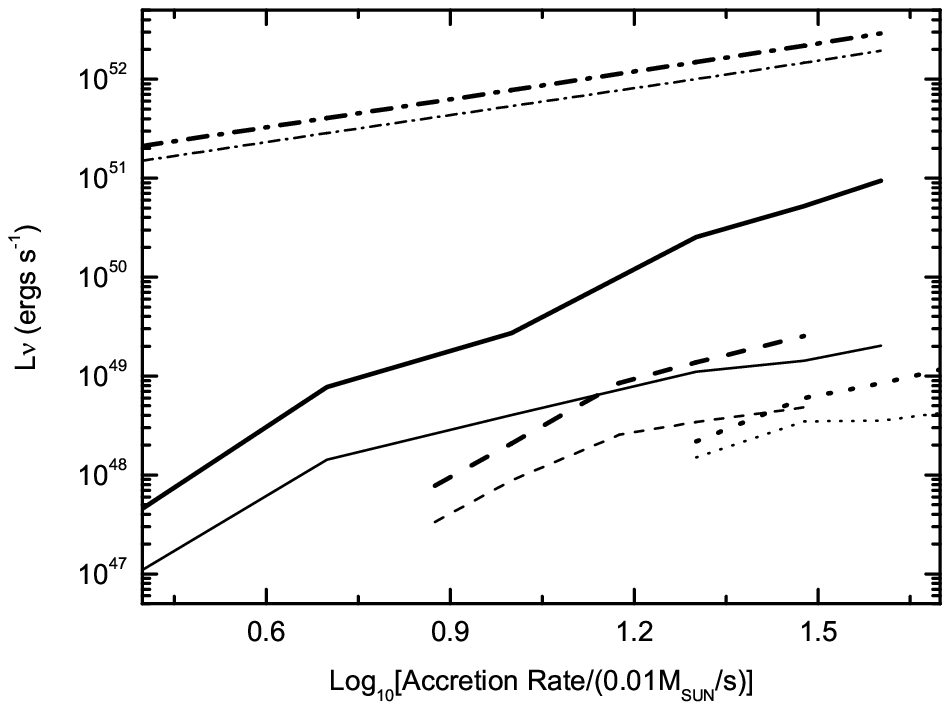}}
\caption{Neutrino annihilation luminosity $L_{\nu\bar{\nu}}$ of the
neutron-star disk with outflow index $s$=0.2 ({\em solid line}), 0.6
({\em dashed line}), 1.0 ({\em dotted line}) and the maximum value
of total luminosity $L_{\nu}$ in the case of $s$=0.2 ({\em
dash-dotted line}), where we take $\varepsilon$=0.8. Left panel
shows the results of model O1 and right panel to model O2. The thin
lines correspond to the luminosity to $\eta_{s}=0$ and thick lines
to $\eta_{s}=0.5$.}
\end{figure}

\newpage
\begin{figure}
\resizebox{\hsize}{!} {\includegraphics{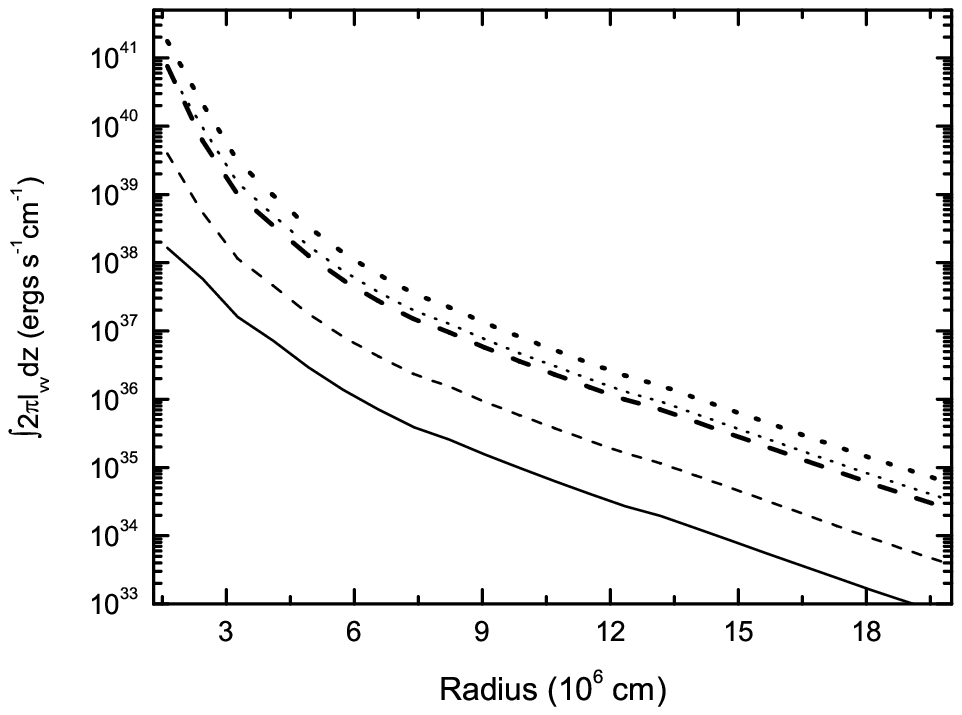}
\includegraphics{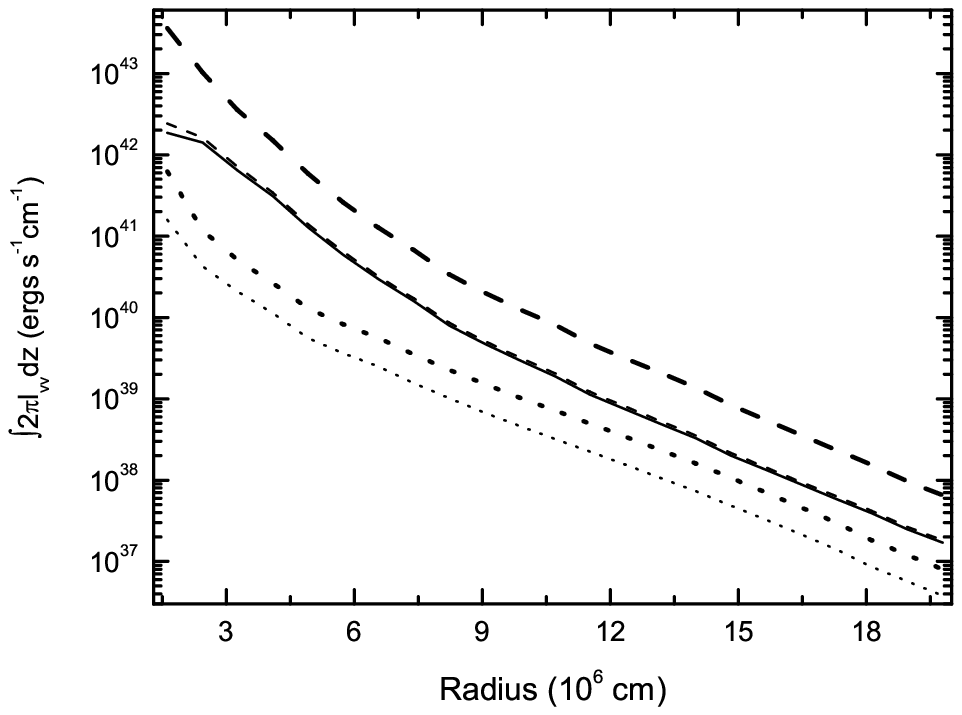}}
\caption{Neutrino annihilation luminosity per cm distribution as a
function of radius for $\dot{M}$=0.01$M_{\odot}$ s$^{-1}$ ({\em left
panel}) and $\dot{M}$=0.1$M_{\odot}$ s$^{-1}$ ({\em right panel}).
(a) {\em Left panel}: annihilation luminosity per cm for a
black-hole disk ({\em thin solid line}), neutron-star disks with
entropy-conservation inner disk and $\eta_{s}$=0 ({\em thin dashed
line}), and $\eta_{s}$=0.5 ({\em thick dashed line}), and
neutron-star disks with advection-dominated inner disk
$\varepsilon$=0.2 and $\eta_{s}$=0 ({\em thin dotted line}), and
$\varepsilon$=0.2 and $\eta_{s}$=0.5 ({\em thick dotted line}). (b)
{\em Right panel}: annihilation luminosity per cm for a neutron-star
disk with outflow (model O2) $\varepsilon$=1, $s$=0.6, $\eta_{s}$=0
({\em thin dotted line}), $\varepsilon$=1, $s$=0.6, $\eta_{s}$=0.5
({\em thick dotted line}). The thin solid line, thin dashed line and
thick dashed line are the same as the left panel.}
\end{figure}

\newpage
\begin{figure}
\resizebox{\hsize}{!} {\includegraphics{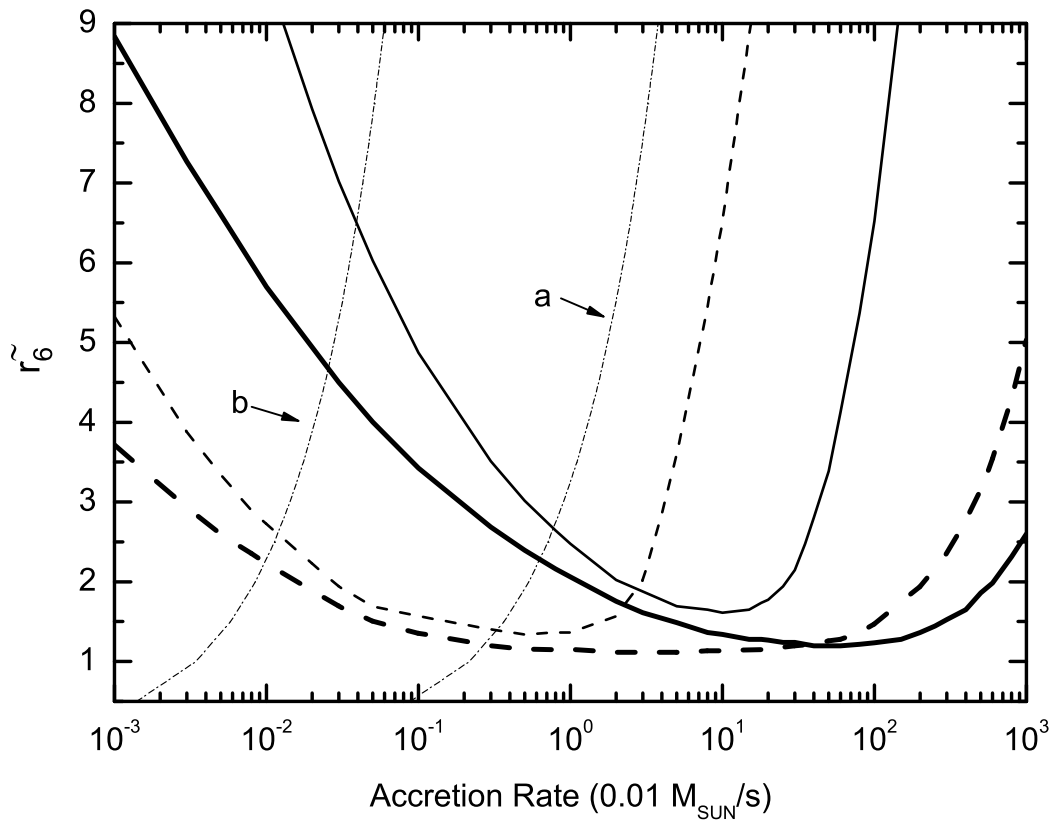}}\caption{The
radius $\tilde{r}$ (in units of $10^{6}$ cm) between the inner and
outer disks with the viscosity parameter $\alpha$=0.1 ({\em solid
line}) and 0.01 ({\em dashed line}). The thick lines show the
results based on the consideration $\Omega\simeq$ const in the inner
disk in \S 5, and the thin lines based on the angular velocity
distribution $\Omega\propto r^{-3/2}$ as in \S 3. The dashed-dotted
lines ``a" and ``b" correspond to the characteristic curve of
$\beta$-equilibrium as in Fig. 1 with the viscosity parameter
$\alpha$=0.1 and $\alpha$=0.01.}
\end{figure}

\newpage
\begin{figure}
\resizebox{\hsize}{!} {\includegraphics{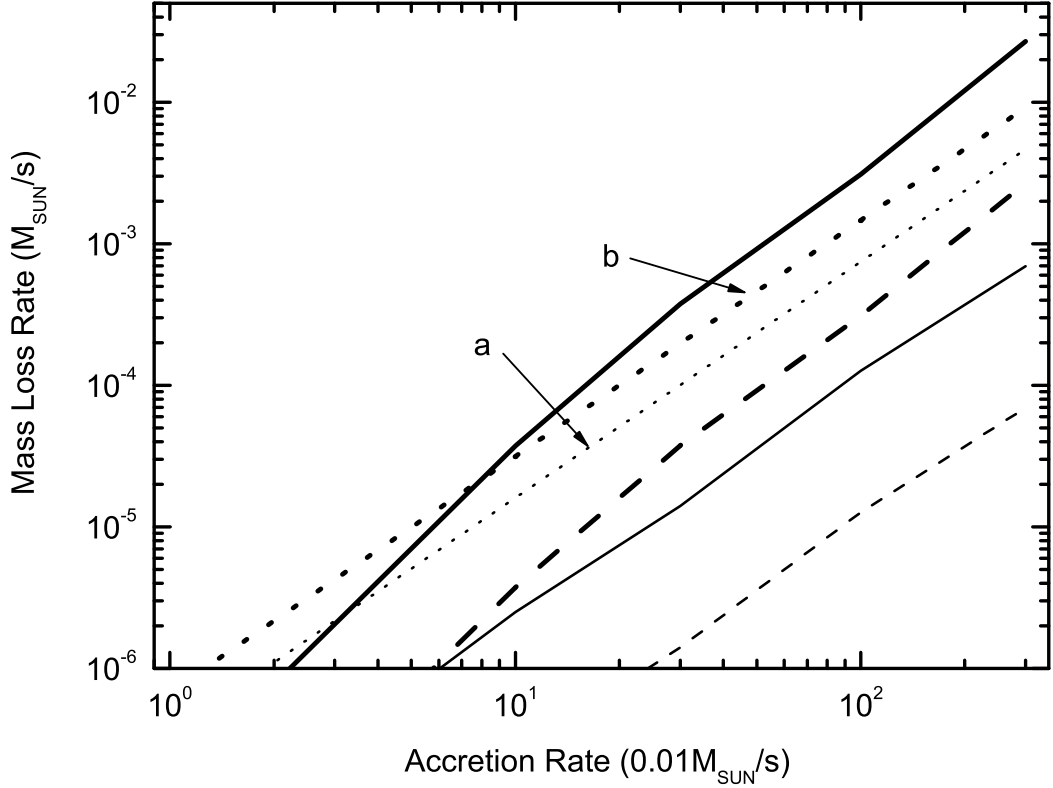}}\caption{Upper
limit of mass-loss rate due to neutrino-driven wind from the neutron
star surface for the outflow being accelerated to $\Gamma=10$
(moderately relativistic) with the surface emission boundary
condition $\eta_{s}=0$ (\textit{thin solid line}) or $\eta_{s}=0.5$
(\textit{thin dashed line}); or $\Gamma=100$ (ultrarelativistic)
with the surface emission boundary condition $\eta_{s}=0$
(\textit{thick solid line}) or $\eta_{s}=0.5$ (\textit{thick dashed
line}). The dotted lines ``a" and ``b" correspond to the strength of
a thermally neutrino-driven wind from the stellar surface for the
boundary condition $\eta_{s}=0$ and $\eta_{s}=0.5$ respectively..}
\end{figure}
\end{document}